\documentclass[%
 reprint,
 amsmath,amssymb,
 aps,
]{revtex4-2}

\usepackage{graphicx}% Include figure files
\usepackage{dcolumn}% Align table columns on decimal point
\usepackage{bm}% bold math
\usepackage[dvipsnames]{xcolor}
\usepackage{hyperref}

\newcommand{\eps}{\varepsilon_{\mathrm{r}}\varepsilon_0}
\newcommand{\dd}{\mathrm{d}}
\newcommand{\rb}{\mathbf{r}}
\newcommand{\kb}{\mathbf{k}}

\newcommand{\F}{\mathcal{F}}

\newcommand{\Om}{\Omega[\{\rho\}]}
\newcommand{\rhos}{\{\rho\}}
\usepackage{ulem}

\definecolor{green}{rgb}{0.1, 0.5, 0.2}

\begin{document}

\preprint{APS/123-QED}

\title{In-plane structure of the electric double layer in the primitive model using classical density functional theory}% Force line breaks with \\

\author{Peter Cats}
 \affiliation{Institute of Physics, University of Freiburg, Hermann-Herder-Str. 3, 79104 Freiburg, Germany}\email{Corresponding Author: andreas.haertel@physik.uni-freiburg.de}
\author{Andreas H\"artel}
\affiliation{Institute of Physics, University of Freiburg, Hermann-Herder-Str. 3, 79104 Freiburg, Germany}

\date{\today}% It is always \today, today,

\begin{abstract}
The electric double layer (EDL) has a pivotal role 
in screening charges on surfaces as in supercapacitor electrodes 
or colloidal and polymer solutions. Its structure is determined 
by correlations between the finite-sized ionic charge carriers 
of the underlying electrolyte and, this way, these correlations 
affect the properties of the EDL and of  applications utilizing EDLs. 
We study the structure of EDLs within classical density 
functional theory (DFT) in order to uncover whether a structural 
transition in the first layer of the EDL that is driven by 
changes in the surface potential depends on specific particle interactions or has a general footing. This transition has been 
found in full-atom simulations.
Thus far, investigating the in-plane structure of the EDL for the primitive model (PM) using DFT proved a challenge. We show here that the use of an appropriate functional predicts the 
in-plane structure of EDLs in excellent agreement with 
molecular dynamics (MD) simulations. 
This provides the playground to investigate how the structure 
factor within a layer parallel to a charged surface changes as 
function of both the applied surface potential and its separation from the surface. 
We discuss pitfalls in properly defining an in-plane structure factor and fully map out the structure of the EDL within the PM for a wide range of electrostatic electrode potentials.
However, we do not find any signature of a structural crossover and conclude that the previously reported effect is not fundamental but rather occurs due to the specific force field of ions used in the simulations.
\end{abstract}

\maketitle

%\tableofcontents
\section{Introduction}
Electrolytes can be found almost anywhere, and attract therefore a lot of interest across a wide variety of disciplines \cite{simon_nm7_2008,porada_pms58_2013,haertel_ees8_2015,verwey_book_1948,kornyshev_rmp79_2007,shapiro_nc3_2012}. 
% Supercapacitors: simon_nm7_2008
% Water desalination: porada_pms58_2013
% Thermal energy harvesting: haertel_ees8_2015
% Chemical and colloidal interactions: verwey_book_1948
% DNA: kornyshev_rmp79_2007
% Nervous conduction: shapiro_nc3_2012
They have been studied for more than a century \cite{Helmholtz,DH,Gouy,Chapman}, but there is still uncharted territory to discover %\cite{Bultmann_2022,Pelagejcev_2022}. 
\cite{jones_arxiv_2022,haertel_prl_2023,Erne2022}. 
% Ion dynamics: jones_arxiv_2022
% Underscreening: haertel_prl_2023
% Heat in EDLs: Erne2022
A particular focus lays on investigating the electric double layer (EDL), \textit{i.e.} the electrode-electrolyte interface, where surface charges are screened by mobile ionic charges from the electrolyte. 
This ability of ions to screen each other is closely related to structure formation in the EDL that is driven by an interplay between electrostatic and steric particle interactions \cite{Gebbie_etal_2015,Smith_etal_2016}. 
Thus, understanding EDLs implies understanding their structure, which we study in this work.
% REVIEW: 
%Structure of electric double layers in capacitive systems and to %what extent (classical) density functional theory describes it.
%A. Härtel.
%J. Phys.: Condens. Matter 29, 423002(24pp) (2017)
%DOI: 10.1088/1361-648X/aa8342 

The structure of a liquid electrolyte can be seen from its particle's and charge's distributions. These density profiles measure the local density and, in the electrolyte's bulk, they do not show any structure, thus, they are constant. This changes in the vicinity of a flat electrode, where the flat boundary induces layering and ordering of particles and charges. 
Here, density profiles are typically resolved perpendicular to the flat electrode, reflecting translational symmetry along the wall. Classical density functional theory (DFT) predicts these perpendicular density profiles of ions very well \cite{Roth_2016,Haertel_2017,Cats_2020_decay,Cats_2020,Cats_2021,Bultmann_2022}. In this manuscript we take another perspective and focus on the in-plane structure of the EDL, \textit{i.e.} the distribution of ions parallel to the interface. 

Merlet and co-workers \cite{Merlet_2014} studied the in-plane structure of the EDL for a model capacitor consisting of liquid butylmethylimidazolium hexafluorophosphate and graphite electrodes. They performed molecular dynamics (MD) simulations using sophisticated force fields and found hexagonal in-plane ordering for certain electrode potentials by considering an "in-plane" structure factor. We will discuss possible pitfalls in introducing a well-defined in-plane structure factor in this work. To clarify whether this ordering  effect is a consequence of the  specific force field  or a more fundamental property that also occurs in the primitive model of electrolytes, the same system has been mapped onto the primitive model a few years later  and the in-plane structure has been studied by employing both DFT and MD  \cite{Hartel_2016}.  For their DFT study, however, the authors used a rather simple mean-field description that did not perform well across almost all parameters considered. The question, therefore, remains whether  the observed ordering effect can be found in the primitive model of electrolytes as well and whether  DFT, when using a  more sophisticated approach, can  accurately predict the in-plane structure of the EDL in the primitive model. These questions will be answered shortly.  

% \section{The Structure Factor}\label{sec:S}
% \section{Modelling the System}\label{sec:system_model}
% \section{Success of our Approach}\label{sec:sys1}
% \section{The Structure Changes with the Surface Potential}\label{sec:structure-results}
% \section{Conclusion}\label{sec:conc}
%
% \section{MSAc in the Plane}
% \section{The in-plane Structure Factor Moving through the System}\label{sec:sys2}
In this work, we first precisely define an in-plane structure factor in section~\ref{sec:S}. In section~\ref{sec:system_model}, then, we introduce the physical systems of interest and the model we use. We consider the same system as studied in Ref.~\cite{Hartel_2016}, described in section~\ref{sec:system_model}, but we now consider a DFT functional constructed from the mean-spherical approximation (MSA) that has been proven to work well for primitive model aqueous electrolytes at room temperature \cite{MSAc,Cats_2020_decay,Cats_2021}. Using this functional, we calculate the density profiles and in-plane structure factor for our system and discuss their agreement with results from previous DFT work and MD simulations in section~\ref{sec:sys1}. In section~\ref{sec:structure-results} we present structure factors across the whole EDL for a wide region of electrode potentials and discuss structural changes that depend on the potential, as found previously~\cite{Merlet_2014}. We conclude in section~\ref{sec:conc}.

\section{The Structure Factor}\label{sec:S}

The structure factor for a multi-component inhomogeneous system can be defined as 
$S_{ij}(\vec{k})=\langle\tfrac{1}{N}\rho_{\kb}^{i}\rho_{-\kb}^j\rangle$
with the Fourier components 
$\rho_{\kb}^i=\sum_{n=1}^{N_i}\exp(-i\kb\cdot\rb_n)$ of the microscopic density. 
Equivalently, it can be written in the form \cite{HansenMcDonald2013}
\begin{align}\label{Eq:S_def}
    S_{ij}&(\kb) = \frac{N_i}{N}\delta_{ij}+\\
    &\frac{1}{N}\int\dd\rb \int \dd\rb'\rho_i(\rb)h_{ij}(\rb,\rb') \rho_j(\rb')\exp(-i\kb \cdot (\rb-\rb')),\nonumber
\end{align}
where $N_i$ is the number of particles of species $i$ in the system, $N=\sum_jN_j$ is the total number of particles, $\rho_j(\rb)$ is the density profile of species $j$ as function of the position $\rb$, and $h_{ij}(\rb,\rb')$ is the total correlation function between particles of species $i$ and $j$ located at positions $\rb$ and $\rb'$, respectively. 
The structure factor depends on a wave vector $\kb$ that is the scattering vector if $S$ describes the scattering of a beam.
In the bulk with constant bulk density 
$\rho_i(\rb)\equiv\rho_i=N_i/V$ in a volume $V$,  Eq.~\eqref{Eq:S_def} reduces to
\begin{align}
    S_{ij}(k) = \frac{\rho_i}{\rho_{\mathrm{tot}}}\left[\delta_{ij} + \rho_j\int \dd\rb \,h_{ij}(r)e^{-i\kb\cdot\rb}\right],
    \label{eq:structure-factor-bulk-spherical}
\end{align}
where $\rho_{\mathrm{tot}}=\sum_i\rho_i$, $r=|\rb|$, and $k=|\kb|$ is the radial wave number in spherical coordinates. 

As we aim to study in-plane structure in the presence of a flat wall, it is useful to adopt Eq.~\ref{eq:structure-factor-bulk-spherical} to a cylindrical geometry with
 coordinates $(u,\theta,z)$ that respect the presence of a wall perpendicular to the $z$ direction. 
With $q$ denoting the radial wave number corresponding to the radial in-plane coordinate $u$ in cylindrical coordinates 
and $k_z$ the wave number in the $z$-direction, 
Eq.~\ref{eq:structure-factor-bulk-spherical} can equivalently be written as
\begin{align}
    S_{ij}(q,k_z) &= \frac{\rho_i}{\rho_{\mathrm{tot}}}\left[\delta_{ij} + \rho_j\int_{-\infty}^{\infty} \dd z \,\hat{h}_{ij}(z,q) e^{-ik_z z}\right]\label{Eq:S_cyl},
\end{align}
where we defined the Hankel-transformed total correlation function
\begin{align}
    \hat{h}_{ij}(z,q) = \int_0^{2\pi}\dd\theta\int_0^\infty\dd u \, u\, h_{ij}(u,z)e^{-iqu\cos\theta} , 
\end{align} 
which is a Fourier-transformed version of the total correlation function $h_{ij}(\rb)$ in only two dimensions.

In a homogeneous isotropic bulk system, one can argue that
\begin{align}
    S_{ij}(k) &= S_{ij}(q=k,k_z=0) \equiv S^*_{ij}(q);\label{Eq:S_symm}\\
    h_{ij}(r) &= h_{ij}(u=r,z=0)\equiv h_{ij}^*(u)\label{Eq:h_symm},
\end{align}
\textit{i.e.} it does not matter in which direction one is looking.
However, one can use these relations and combine Eqs.~\eqref{Eq:S_symm} and~\eqref{Eq:S_cyl} to find
\begin{align}
    S^*_{ij}(q) = \frac{\rho_i}{\rho_{\mathrm{tot}}}\left[\delta_{ij} + \rho_j\int_{-\infty}^{\infty} \dd z \,\hat{h}_{ij}(z,q)\right]
    \label{eq:structure-factor-bulk-q-direction}
\end{align}
in bulk.

\subsection{Structure Factor in the Plane}
Let us now confine the integration limits in Eq.~\eqref{eq:structure-factor-bulk-q-direction} and consider a slab of thickness $L'$ in the $z$-direction around (any) $z_0$ in the bulk, \textit{i.e.} $z\in [z_0-L'/2,z_0+L'/2]$. Then Eq.~\ref{eq:structure-factor-bulk-q-direction} becomes
\begin{align}\label{Eq:S_slab_bulk}
    S_{ij}^*(q;L') = \frac{\rho_i}{\rho_{\mathrm{tot}}}\left[\delta_{ij} + \rho_j\int_{-L'/2}^{L'/2} \dd z\, \hat{h}_{ij}(z+z_0,q)\right].
\end{align} 
This equation represents the in-plane structure factor calculated only within the slab. Clearly, in the limit 
\begin{align}\label{Eq:S_lim_bulk}
    \lim_{L'\rightarrow\infty} S_{ij}^*(q;L') = S_{ij}(k),
\end{align}
\textit{i.e.} $S_{ij}^*(q;L')$ converges to the bulk structure factor $S_{ij}(k)$, as expected.
In a similar fashion, we can confine the integration limits in  Eq.~\eqref{Eq:S_def}. For this purpose, we first restrict the particle number to the same slab and replace $N_i$ and $N$ by $n_i$ and $n_{\mathrm{tot}}$, respectively, where
\begin{align}
   n_i(z_0,L') = \int_{z_0-L'/2}^{z_0+L'/2}\dd z \,\rho_i(z)
\end{align}
is the number of particles of species $i$ in the slab of thickness $L'$ centered around $z_0$ per unit area, and ${n_{\mathrm{tot}}(z_0,L')=\sum_in_i(z_0,L')}$.
Then, confining the integration limits in the $z$-direction, 
the in-plane structure factor for an inhomogeneous system (still homogeneous in $x$ and $y$ directions) reads
\begin{widetext}
\begin{align}\label{Eq:S_3Dip}
    S_{ij}^*(q,k_z;z_0,L') = \frac{n_i(z_0,L')}{n_{\mathrm{tot}}(z_0,L')}\delta_{ij}+\frac{1}{n_{\mathrm{tot}}(z_0,L')}\int_{z_0-L'/2}^{z_0+L'/2} \dd z\int_{z_0-L'/2}^{z_0+L'/2} \dd z' \rho_i(z)\hat{h}_{ij}(z,z',q)\rho_j(z')e^{-ik_z(z-z')},
\end{align}
\end{widetext}
where $\hat{h}_{ij}(z,z',q)$ is the inhomogeneous form of $\hat{h}_{ij}(z,q)$, because in bulk $\hat{h}(z,z',q)=\hat{h}(z-z',q)\equiv \hat{h}(z,q)$.
Important to notice is that Eq.~\eqref{Eq:S_3Dip} is the three-dimensional structure  factor determined only in a finite slab of thickness $L'$ around $z_0$. Taking the limit of vanishing $L'$ causes the integral to vanish, \textit{i.e.} the integral is naturally dependent on the integration limits. Naturally, taking the limit of $L'$ to infinity returns Eq.~\eqref{Eq:S_def}.  

In order to make practical use of Eq.~\eqref{Eq:S_3Dip}, we first need to simplify it. 
Motivated by Eq.~\eqref{Eq:S_symm} and~\eqref{Eq:S_slab_bulk}, we further study the case in which one sets $k_z=0$. We have in mind the idea of a test particle placed at $z_0$ in the center of the slab. Accordingly, we replace the first density profile in the integrand by $\rho_i(z) = n_i(z_0,L')\delta(z-z_0)$. By doing so, we can reduce Eq.~\eqref{Eq:S_3Dip} to
\begin{align}\label{Eq:S_ipin}
    S^*_{ij}(q;z_0,L') = & \frac{n_i(z_0,L')}{n_{\mathrm{tot}}(z_0,L')}H_{ij}(q;z_0,L');\\
    H_{ij}(q;z_0,L') = &\delta_{ij}+\int_{z_0-L'/2}^{z_0+L'/2}\dd z \, \hat{h}_{ij}(z_0,z,q)\rho_j(z),\label{Eq:H}
\end{align}
which in the limit $z_0\rightarrow\infty$ (in bulk) is identical to Eq.~\eqref{Eq:S_slab_bulk}. The quantity $H_{ij}$, the normalized structure factor, is introduced for convenience, which will become clear in section~\ref{sec:sys1} and in appendix~\ref{sec:sys2}. Note that Eq.~\eqref{Eq:S_ipin} is an approximation of Eq.~\eqref{Eq:S_3Dip}, made for practical purposes.

\subsection{The Total Correlation Function in DFT}

In order to calculate any of the previous in-plane structure factors, one has to have access to the total correlation function $\hat{h}_{ij}(z,z',q)$. This quantity can be determined via the Hankel-transformed Ornstein-Zernike (OZ) equation~\cite{Hartel_2015_2,Hartel_2016,Tschopp_2021}
\begin{align}\label{Eq:OZ_hankel}
    \hat{h}_{ij}(z,z',q)& = \hat{c}^{(2)}_{ij}(z,z',q) + \\
    &\sum_k \int \dd z'' \hat{h}_{ik}(z,z'',q) \rho_k(z'') \hat{c}^{(2)}_{kj}(z'',z',q)\nonumber,
\end{align}
in which the Hankel transform is defined by
\begin{align}
    \hat{f}(q) & = \int_0^\infty \dd u \, u \int_0^{2\pi} \dd \theta f(u) e^{iqu\cos\theta}\\
    & = 2\pi \int_0^\infty \dd u \, u J_0(qu) f(u),
\end{align}
with $J_0(qu)$ the zeroth-order Bessel function of the first kind. Eq.~\eqref{Eq:OZ_hankel} can be solved iteratively via Picard iterations. 

The question whether we can detect certain (crystalline-like) in-plane structures is already answered in Eq.~\eqref{Eq:OZ_hankel}, because, although we have an inhomogeneous system on the $z$-axis, we assume translational and rotational symmetry in the plane perpendicular to the z-axis. Hence, with this approach, one is not able to {find a respective order} in the $xy$-plane. However, the approach gives access to structure in the $xy$-plane and, thus, would allow for detecting a precursor of an ordering transition, similar to the total correlation structure predicted by the OZ equation in a homogeneous bulk system: For hard-disk systems it is known that there is a signature in the second peak of the (``homogeneous bulk'') total correlation function close to the freezing point~\cite{Truskett_1998,Ramakrishnan_1979,Gonzales_1991}. In essence, when we restrict our study to only the first layer next to the surface, we have a hard-disk like system and the question therefore is whether we can observe these subtle signatures in the in-plane structure factor and total correlation function. 

\subsection{Charge-charge and density-density structure}
For convenience, and for future reference, let us define the charge-charge (ZZ) and number-number (NN) structure factors by, respectively, \cite{HansenMcDonald2013}
\begin{align}\label{Eq:S_ZZ}
    S_\textrm{ZZ} = \sum_{ij} Z_i Z_j S_{ij}
    \end{align}
    and
\begin{align}\label{Eq:S_NN}
    S_\textrm{NN} = \sum_{ij} S_{ij},
\end{align}
where the summations are over the number of species.

\section{Modelling the System}\label{sec:system_model}

\begin{figure*}
    \centering
    \includegraphics[width = 0.75\textwidth]{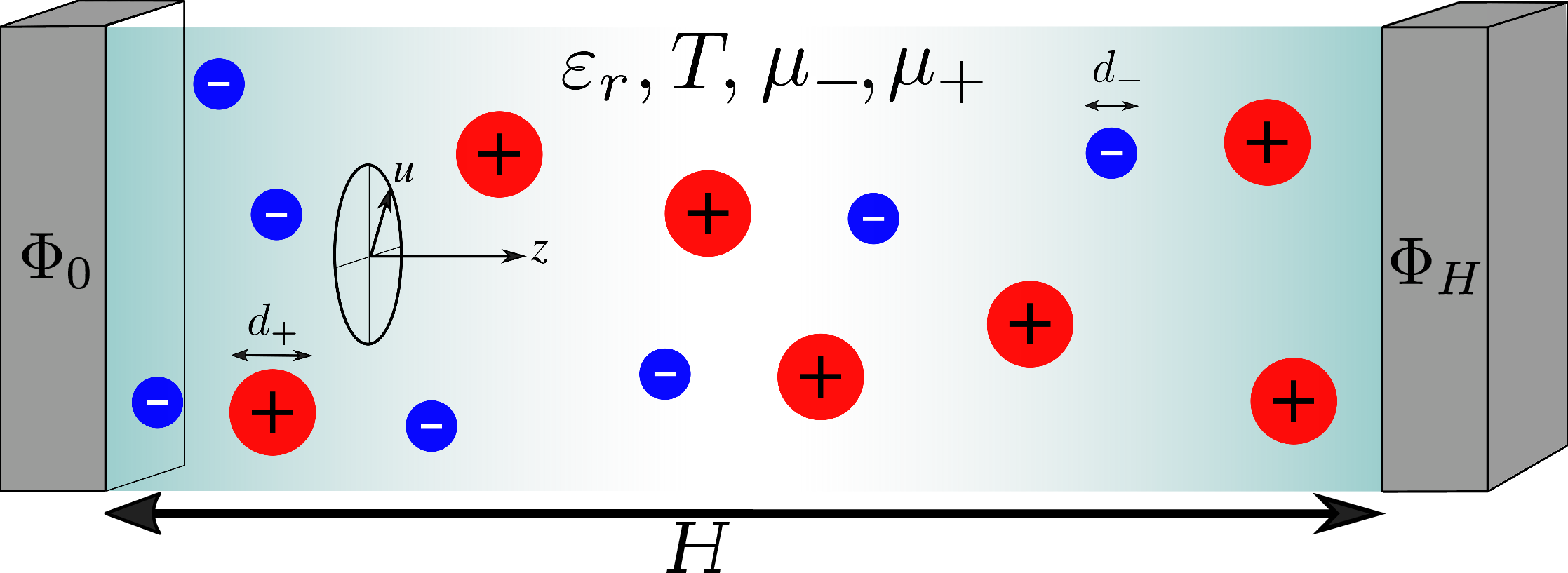}
    \caption{The system under consideration. Two planar hard walls are located at $z=0$ and $z=H$ with surface potentials $\Phi_0$ and $\Phi_H$, respectively. The surfaces confine  an electrolyte at temperature $T$ in chemical equilibrium with a reservoir at chemical potentials $\mu_\pm$ with which it can exchange ions. The electrolyte is modelled in the Primitive Model (PM) with ions as charged hard spheres of diameters $d_\pm$ that carry charges $\pm 0.785$e in their centre. The ions move in a continuum medium, characterized by the relative dielectric permittivity $\varepsilon_r$. }
    \label{Fig:system}
\end{figure*}

The system under consideration is an electrolyte at temperature $T$  that is confined  between two parallel flat hard walls located at $z=0$ and $z=H$ at which electrostatic surface potentials $\Phi_0$ and $\Phi_H$ are applied, respectively, as depicted in Fig.~\ref{Fig:system}. The electrolyte is in chemical equilibrium with an ion reservoir at chemical potentials $\mu_\pm$ (corresponding to bulk densities $\rho_{\textrm{b},\pm}$) for the cations and anions, respectively~\footnote{We consider a grand-canonical ensemble where no chemical reactions are involved.}. 
 The respective non-electrostatic ion-electrode interaction potential between ions of species $i$ and the wall at $z=0$ is
\begin{align}
\beta V_\textrm{ext}^{i}(z) = \begin{cases}
\infty &\quad \text{ for } z \leq d_{i}/2;\\
0 &\quad \text{ for } z > d_{i}/2,\\
\end{cases}
\end{align} 
where $\beta=1/k_\textrm{B}T$ with $k_\textrm{B}$ Boltzmann's constant. The electrostatic part of the ion-electrode interactions is treated via the Poisson equation
\begin{align}
    \partial_z^2\Phi(z) = -\frac{\rho_\textrm{Ztot}(z)}{\eps},
\end{align}
where $\rho_\textrm{Ztot}(z) = \rho_\textrm{Z}(z)+\delta(z)\sigma_0 + \delta(z-H)\sigma_H$ denotes the total charge density in the system, \textit{i.e.} the charge density of the mobile ions  $\rho_\textrm{Z}(z)$ plus the surface charge densities $\sigma_0$ and $\sigma_H$ of the respective walls. Note already that 
we use an external potential that only depends on the out-of-plane coordinate $z$ without variations in the $xy$-plane.

For the electrolyte we use the Primitive Model (PM), in which the ions are modelled as charged hard spheres of diameters $d_j$ residing in a continuum dielectric medium, characterized by the Bjerrum length $\lambda_{\mathrm{B}}=\beta e^2/4\pi\eps$, where $e$ is the proton charge and $\eps$ the dielectric permittivity of the medium. The ion-ion interaction potential for two ions separated by a distance $r$ is then given by
\begin{align}
\beta u_{ij}^\textrm{PM}(r) = \begin{cases}
\infty &\quad \text{ for } r \leq d_{ij}/2;\\
Z_iZ_j\frac{\lambda_{\mathrm{B}}}{r} &\quad \text{ for } r > d_{ij}/2,\\
\end{cases}
\end{align}
where $Z_j$ denotes the valency of the ions of species $j$, and $d_{ij}=(d_i+d_j)/2$ the common diameter of species $i$ and $j$. Specifically, we consider a system with ion diameters $d_-=0.506$ nm and $d_+=0.618$ nm, valencies $Z_\pm=\pm  0.785$, and Bjerrum length $\lambda_\textrm{B}=4.17$ nm ($T=400$ K and $\epsilon_r=10$), mimicking the ionic liquid of Ref.~\onlinecite{Merlet_2014}. The system size we use for most of the DFT calculations is $H=10 d_-$, the only exception being the calculation of the density profiles when comparing to the simulation data for which we consider the system size $H=12.32$ nm, in accordance with Ref.~\cite{Hartel_2016}. Typically, the density profiles in our system decay to bulk values within 5 $d_-$ ($\approx 2.5$ nm), as can be seen from Fig.~\ref{Fig:rho}.

We tackle this model using DFT with the MSAc implementation, which details can be found in our previous works~\cite{Cats_2020_decay,Cats_2020,Cats_2021,Haertel_2017} and are summarized in appendix~\ref{App:MSA}. In short, DFT is an  exact theoretical framework to access the structure and thermodynamics of a given system in an external potential $V_\textrm{ext}^j$ \cite{Evans_1979}. The main object within DFT is the grand potential functional $\Omega[\rhos]$ of the densities $\rhos$, which typically is not known exactly. A main problem in DFT is setting up accurate functionals for specific systems. The MSAc functional reads
\begin{widetext}
\begin{align}
    \Omega[\rhos] = \F_\textrm{id}[\rhos]+\F_\textrm{ex}^\textrm{HS}[\rhos]+\F_\textrm{ex}^\textrm{MFC}[\rhos]+\F_\textrm{ex}^\textrm{MSAc}[\rhos] - A\sum_j \int\dd z \,\rho_j(z)(\mu_j-V_\textrm{ext}^j(z))-A\Phi_0\sigma_0-A\Phi_H\sigma_H, \label{Eq:MSAc-functional}
\end{align}
\end{widetext}
where $\F_\textrm{id}[\rhos]$ is the ideal Helmholtz free energy functional, $\F_\textrm{ex}^\textrm{HS}[\rhos]$ the excess Helmholtz free energy functional that deals with the hard-sphere interactions for which we use the White Bear mark II (WBII) version of fundamental measure theory (FMT) \cite{Roth_2010,Hansen_Goos_2006}, $\F_\textrm{ex}^\textrm{MFC}[\rhos]$ is the mean-field functional for the electrostatic (Coulombic) interactions, and $\F_\textrm{ex}^\textrm{MSAc}[\rhos]$ is the beyond mean-field functional for the electrostatic interactions that makes use of the direct correlation functions of MSA~\cite{Waisman_1970,Waisman_1972_I,Waisman_1972_II,Hiroike,MSAc,Cats_2021}. The surface area $A$ is eventually factored out and does not play a role in the DFT calculations. The grand potential functional is minimized by the equilibrium density profiles $\rho^\textrm{eq}_j(z)$, 
\begin{align}
    \left.\frac{\delta \Om}{\delta \rho(z)}\right|_{\rho_j(z) = \rho^\textrm{eq}_j(z)}=0.
\end{align}
From minimizing the grand potential, one only gains access to $\rho_i(z)$ and no information  to what happens in the plane, because all applied external potentials have translational symmetry and we do not break this symmetry in our calculations. In general, DFT could predict also density profiles that are inhomogeneous in the $xy$ plane but the numerical calculations would be too expensive \cite{haertel_prl2012}. We obtain in-plane structure via the OZ equation as explained in the previous section, for which we need the direct pair-correlation function. This, however, follows from the grand potential function via
\begin{align}
    c^{(2)}_{ij}(\rb,\rb') &= -\beta \frac{\delta^2\F_\textrm{ex}[\rhos]}{\delta\rho_i(\rb),\rho_j(\rb')}\\
    =&c^{(2),\textrm{HS}}_{ij}(\rb,\rb')+c^{(2),\textrm{MFC}}_{ij}(\rb,\rb')+c^{(2),\textrm{MSAc}}_{ij}(\rb,\rb')\nonumber,
\end{align}
which naturally contains a HS contribution, a MFC contribution, and a MSAc contribution. Specifically, we are interested in the Hankel-transformed direct-correlation function (see Eq.~\eqref{Eq:OZ_hankel}). The Hankel-transformed $c^{(2),\textrm{HS}}_{ij}(\rb,\rb')$ for the Rosenfeld-version of FMT is well described in Ref.~\onlinecite{Tschopp_2021} for a single-component system. We straightforwardly generalized that approach for the WBII-version of FMT including a tensor correction \cite{Tarazona_2000} to more accurately incorporate dense multicomponent systems. The MFC contribution is simply given by
\begin{align}
    c^{(2),\textrm{MFC}}_{ij}(z,z',q) = -2\pi\lambda_\textrm{B} Z_iZ_j \frac{e^{-q|z-z'|}}{q},
\end{align}
and for the MSAc contribution we numerically Hankel transformed the MSAc direct correlation function, which explicit form can, for example, be found in Ref.~\onlinecite{Cats_2020}.

To summarize our approach, we construct a grand potential functional for the PM electrolyte confined between two planar hard surfaces at which we apply surface potentials. This gives us both access to the density profiles perpendicular to the surfaces $\rho_j(z)$ as well as the direct correlation function $c_{ij}^{(2)}(z,z',u)$, where $u$ is the in-plane radial component. Both these quantities are used to calculate the total correlation function in Hankel space $\hat{h}_{ij}(z,z',q)$, where $q$ is the radial component of the wave vector in the plane. Consequently, we can calculate the in-plane structure factor $S(q,z_0,L')$ within a slab of thickness $L'$ centered around $z_0$.

\section{Success of our Approach}\label{sec:sys1}
In Ref.~\cite{Hartel_2016}, the authors consider an ionic liquid in the PM but similar to the one studied in Ref.~\cite{Merlet_2014}, where in-plane structure was found in simulations using specific force fields. If in-plane structure would form, then one  would expect this to show up in the "in-plane" structure factor (as demonstrated in Ref.~\cite{Merlet_2014}). For the following discussion, we consider $S_{ij}^*(q;z)$ from Eq.~\eqref{Eq:S_ipin} as the in-plane structure factor and $H_{ij}(q;z)$ from Eq.~\eqref{Eq:H}  as the normalized in-plane structure factor \footnote{Note that Eq.~\eqref{Eq:S_ipin} corresponds to the in-plane structure factor calculated in Ref.~\cite{Hartel_2016}, which they present in their Fig.~6 and~7.}. 

\begin{figure}
    \centering
    \includegraphics[width=\columnwidth]{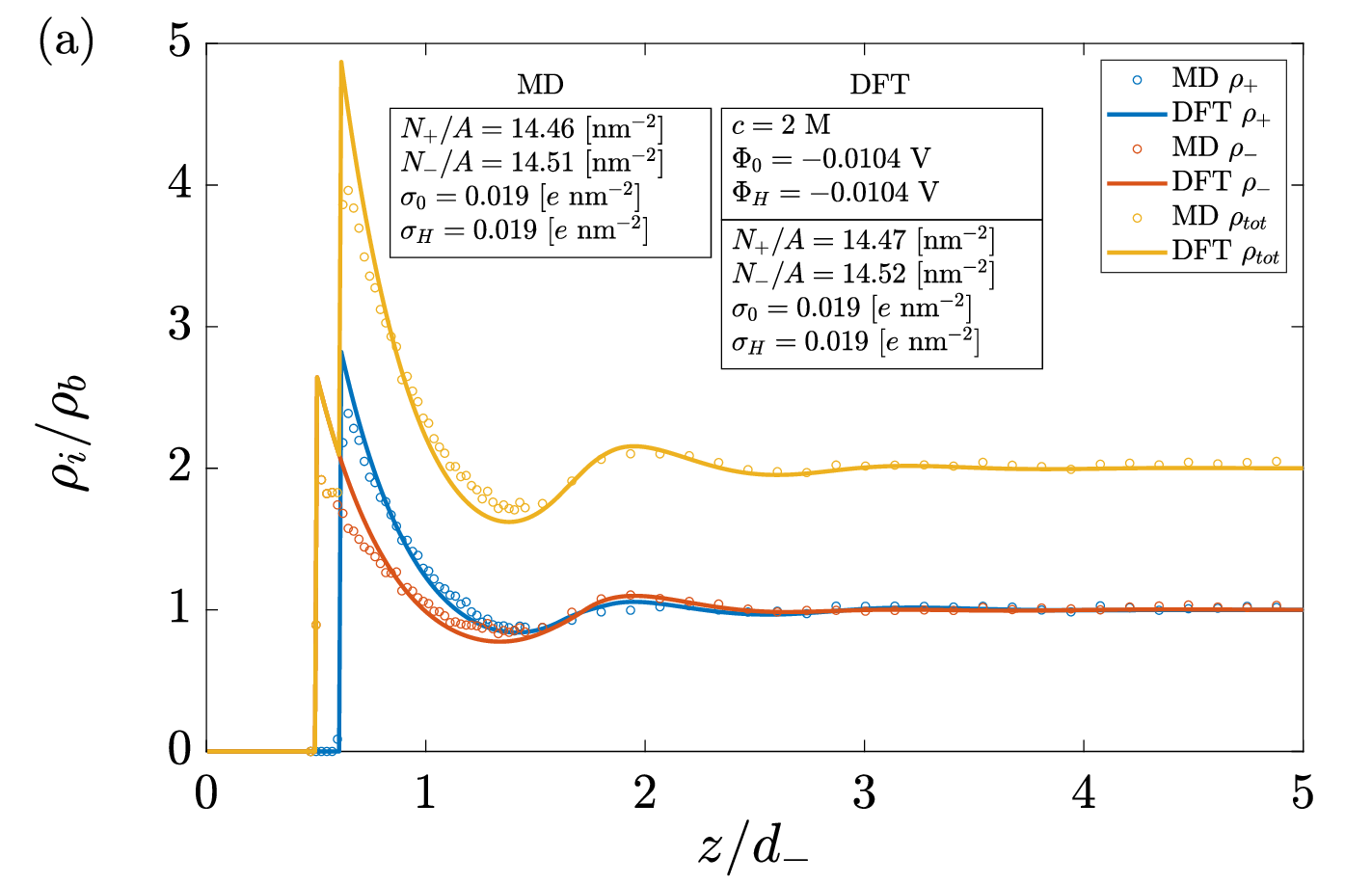}\\
    \includegraphics[width=\columnwidth]{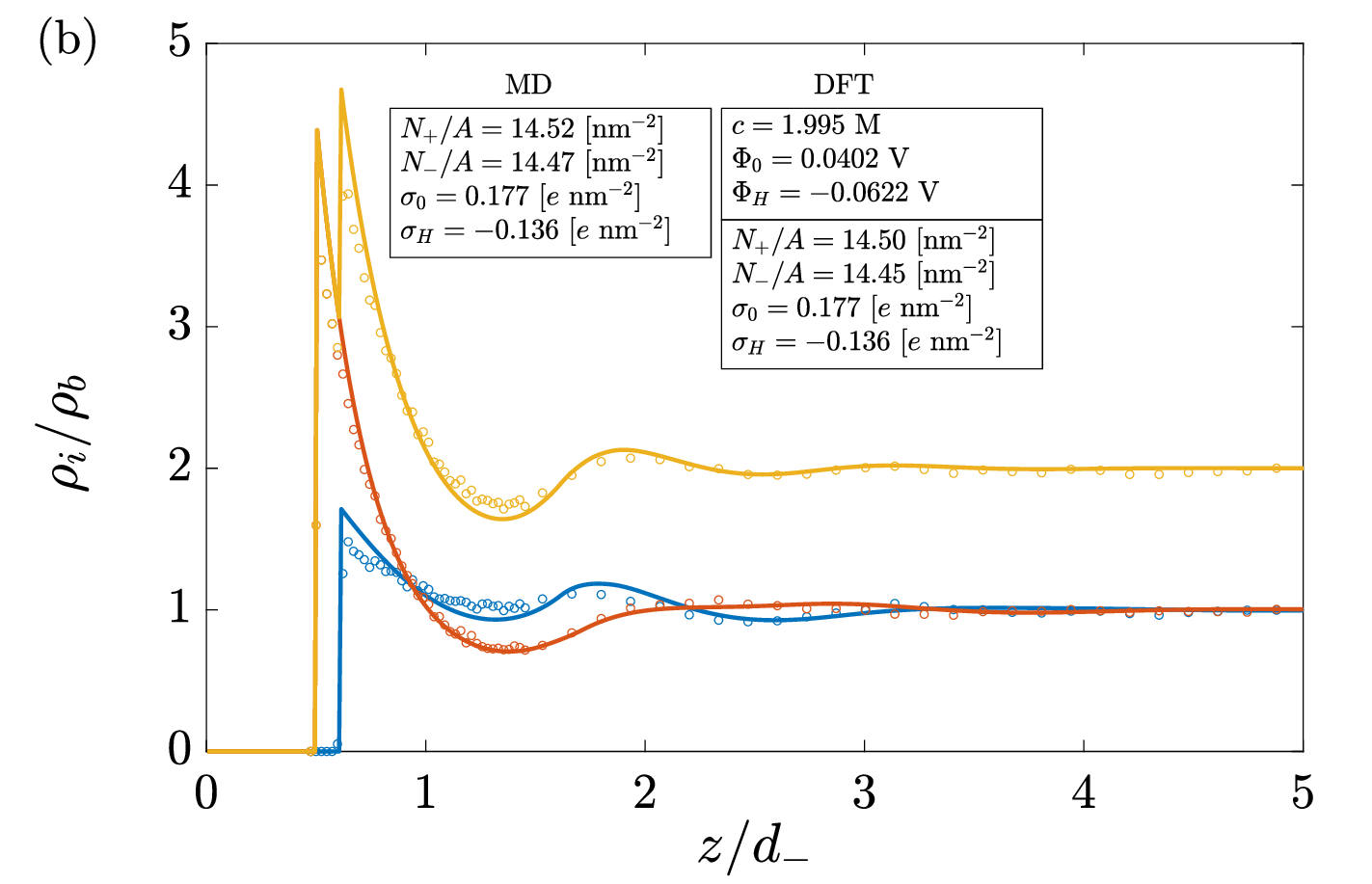}\\
    \includegraphics[width=\columnwidth]{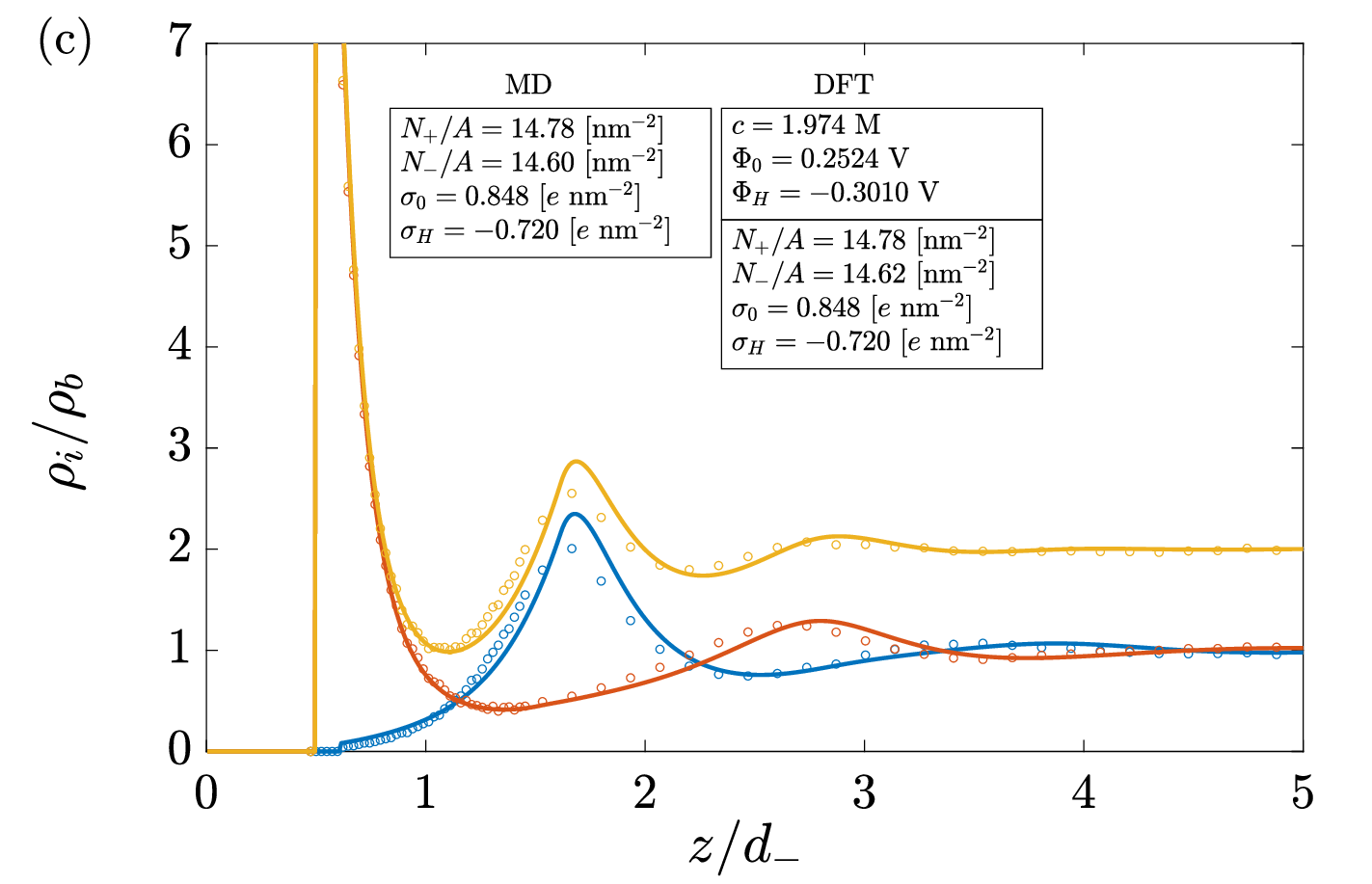}
    \caption{\label{Fig:rho}Density profiles $\rho_i$, normalized to bulk density $\rho_{\textrm{b}}$, for separations $z$ from the wall located at $z_0=0$; $d_{-}$ is the diameter of the negatively charged ions. In blue the cation density profile $\rho_+$ is given, in red the anion density profile $\rho_-$, and in yellow the total density profile $\rho_{\mathrm{tot}}$. The inset in each panel shows the corresponding system parameters, \textit{i.e.}, particle numbers $N_{\pm}$ in the volume $H A$ with wall separation $H$, wall charge densities $\sigma_0$ and $\sigma_{H}$ at the respective walls, and corresponding bulk concentration $c$ and wall potentials $\Phi_0$ and $\Phi_H$ used in the MSAc functional. The surface potential is smallest in (a), increased a bit in (b), and is largest in (c). The Bjerrum length is $\lambda_\textrm{B} = 4.17$ nm.}
\end{figure}

First let us show the density profiles at the positive-charged hard wall and compare those with the MD simulation data given in Ref.~\cite{Hartel_2016}. This comparison is presented in Fig.~\ref{Fig:rho}. Note here that the MD simulations were performed within the canonical ensemble, while the DFT calculations are performed within the grand-canonical ensemble. Hence, we matched the number of particles in the system and the charge on the walls within DFT with those given from the simulations.  The corresponding numbers used as input parameters for DFT (the concentration $c$ and surface potential $\Phi_0$ and $\Phi_H$ at the walls located at $z=0$ and $z=H$) are given within each panel in Fig.~\ref{Fig:rho}. \footnote{Note the big difference in the surface potential from the MFC and MSAc functional that is needed to reproduce the numbers from MD. The surface potentials for the MFC functional are $\Phi_0=0$ V, $\Phi_0 = 0.1$ V, and $\Phi_0=0.5$ V for Fig.~\ref{Fig:rho}(a), Fig.~\ref{Fig:rho}(b), and Fig.~\ref{Fig:rho}(c), respectively.}  Overall, the MD and the MSAc DFT density profiles agree reasonably well.

\begin{figure}
    \centering
    \includegraphics[width=\columnwidth]{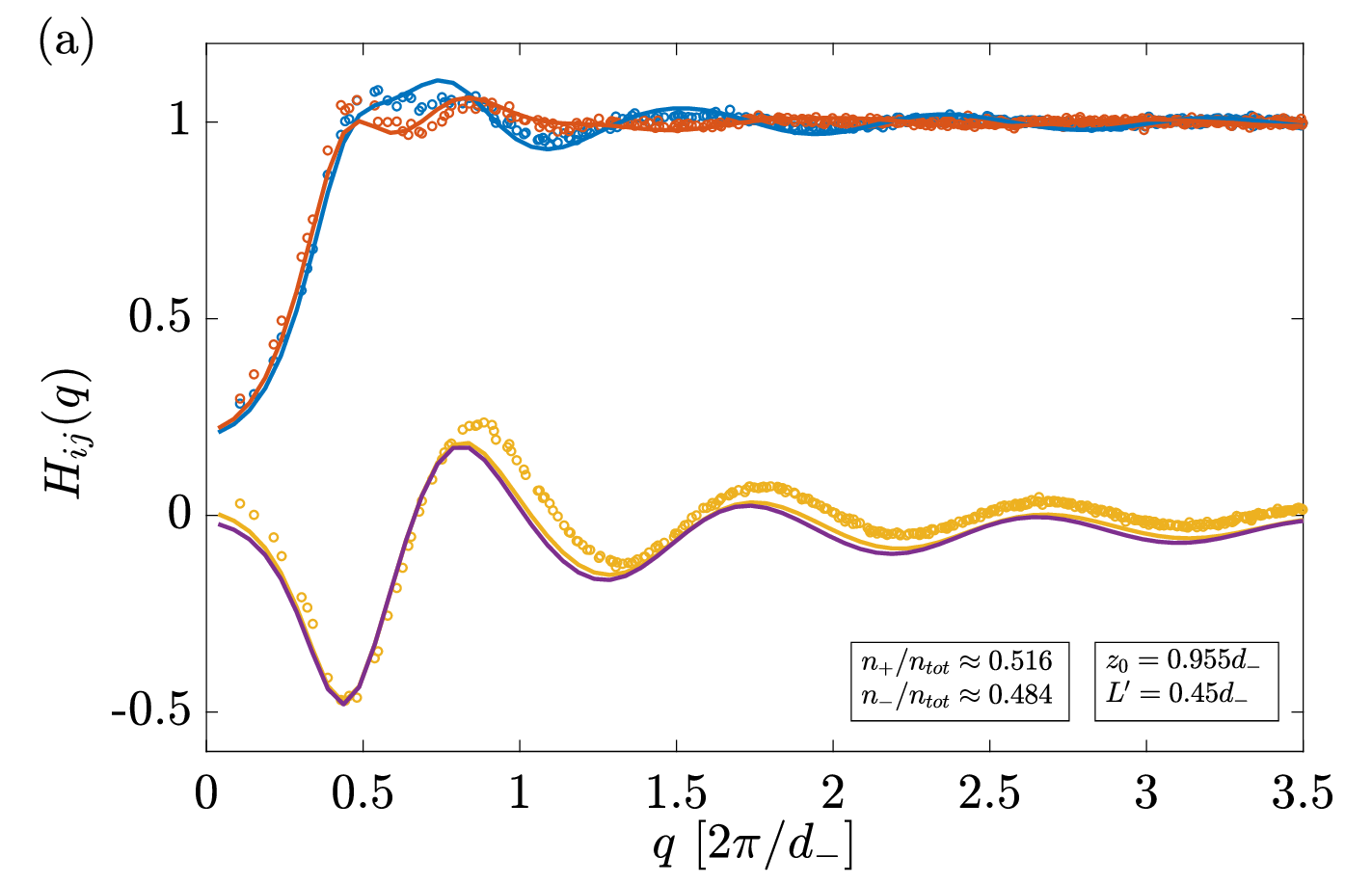}\\
    \includegraphics[width=\columnwidth]{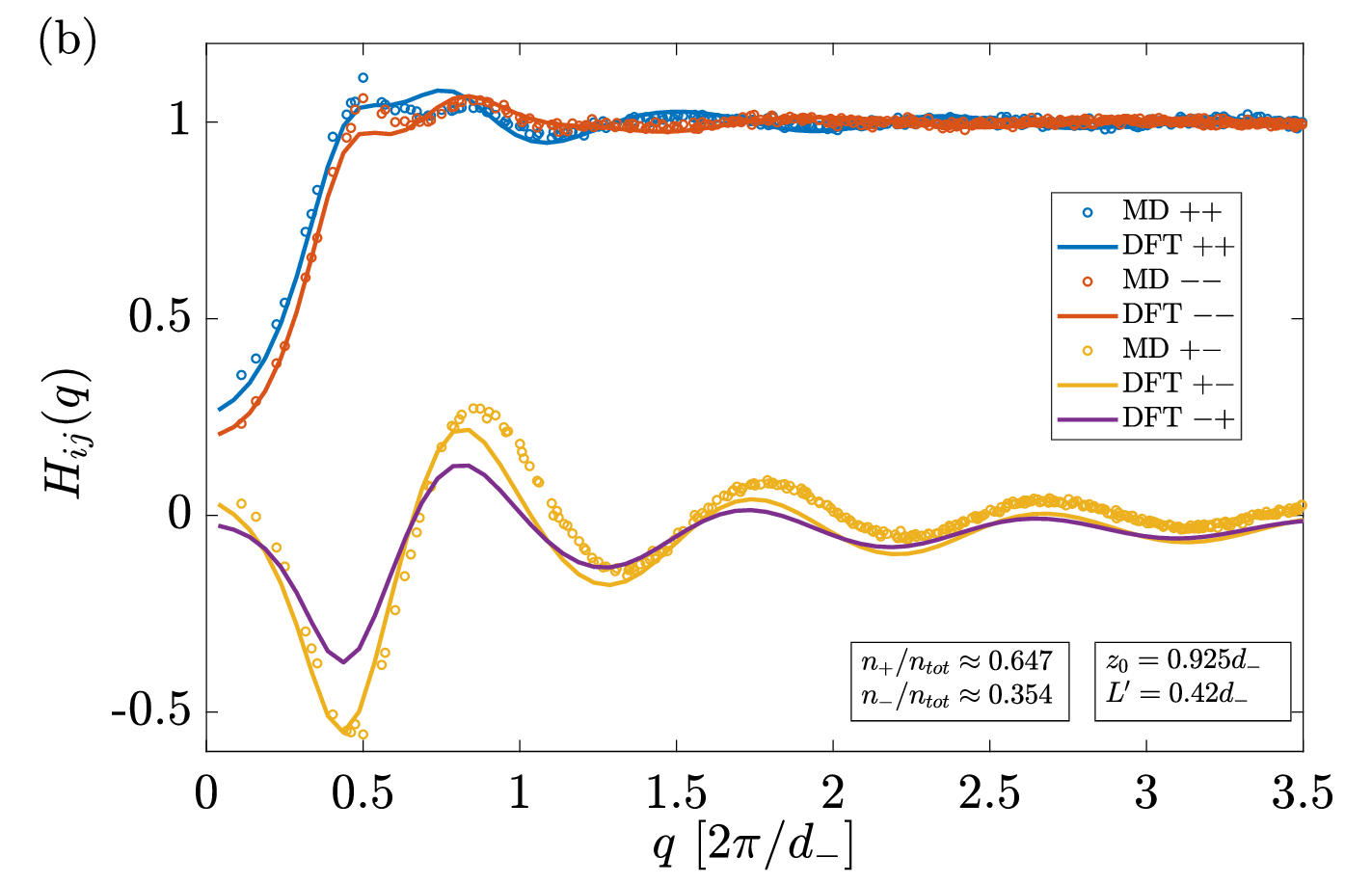}\\
    \includegraphics[width=\columnwidth]{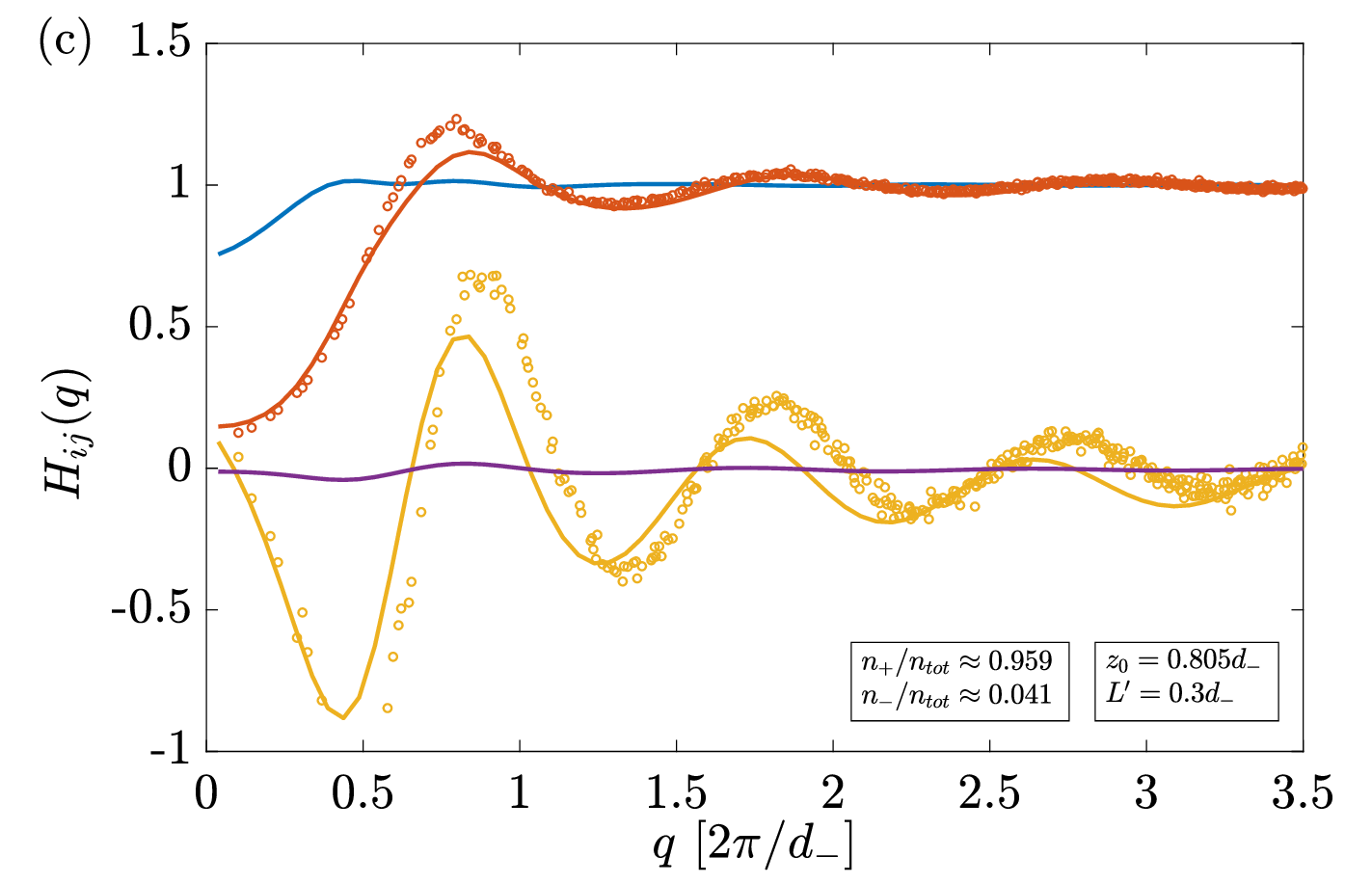}
    \caption{\label{Fig:S}Structure factors $H_{ij}(q)$ from DFT (MSAc functional and MD via Eqs.~\eqref{Eq:H} and ~\eqref{Eq:S_MD}, both for the same parameters and density profiles as given in Fig.~\ref{Fig:rho}; the potential increases from (a) $\Phi_0=-0.0104$ V via (b) $\Phi_0=0.0402$ V to (c) $\Phi_0=0.2524$ V. The values $z_0$ and $L'$ (values according to the numerical grid) are chosen such that $z_0$ lies in between the contact value of the smallest ions (at $d_{-}/2$) and the first minimum (closest to the wall) in the total density profile (see Fig.~\ref{Fig:rho}). In addition, values for the prefactors $n_i(z_0,L')/n_{\mathrm{tot}}(z_0,L')$ are presented (the prefactors appear in Eq.~\eqref{Eq:S_ipin}).}
\end{figure}

\begin{figure}
    \centering
    \includegraphics[width=\columnwidth]{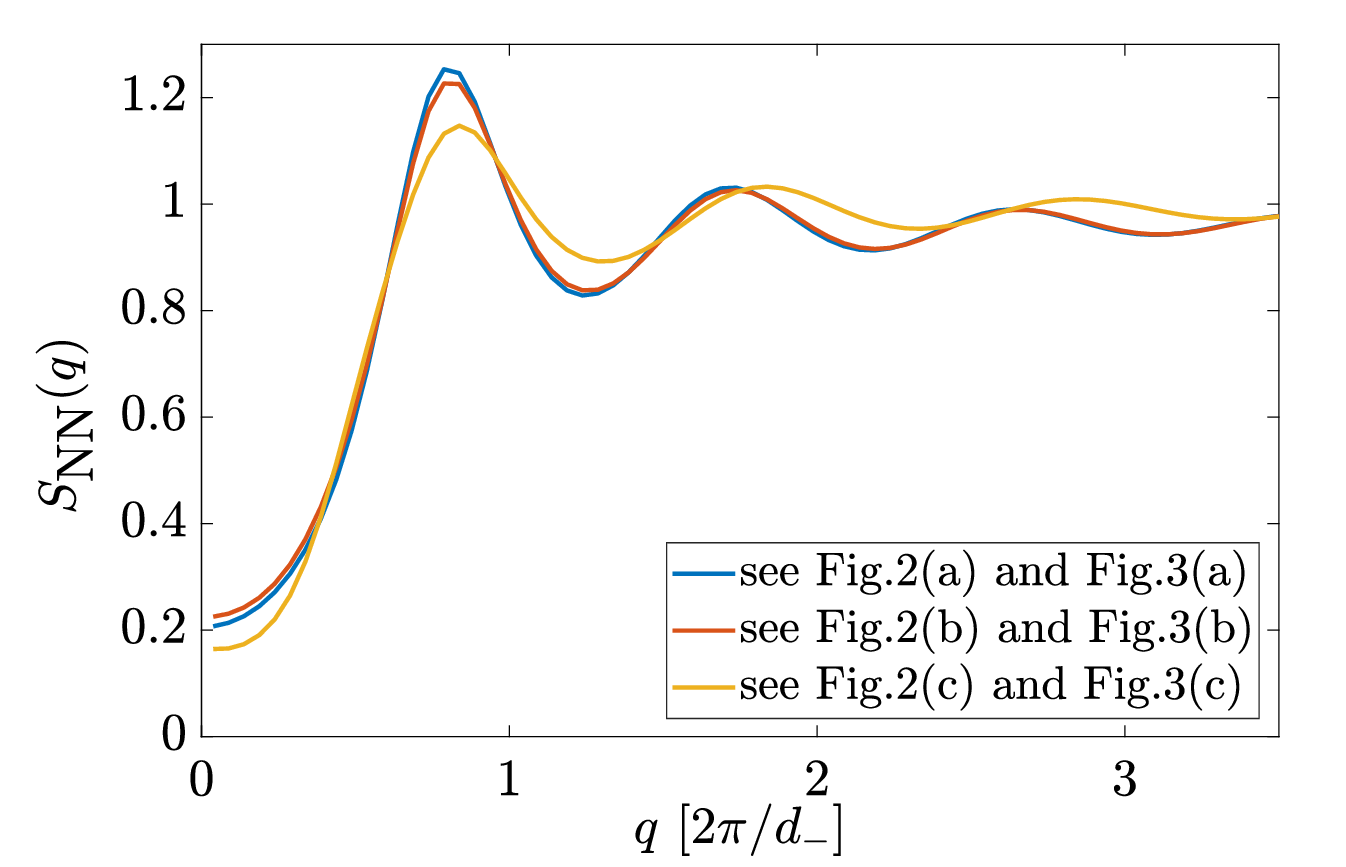}\\
    \includegraphics[width=\columnwidth]{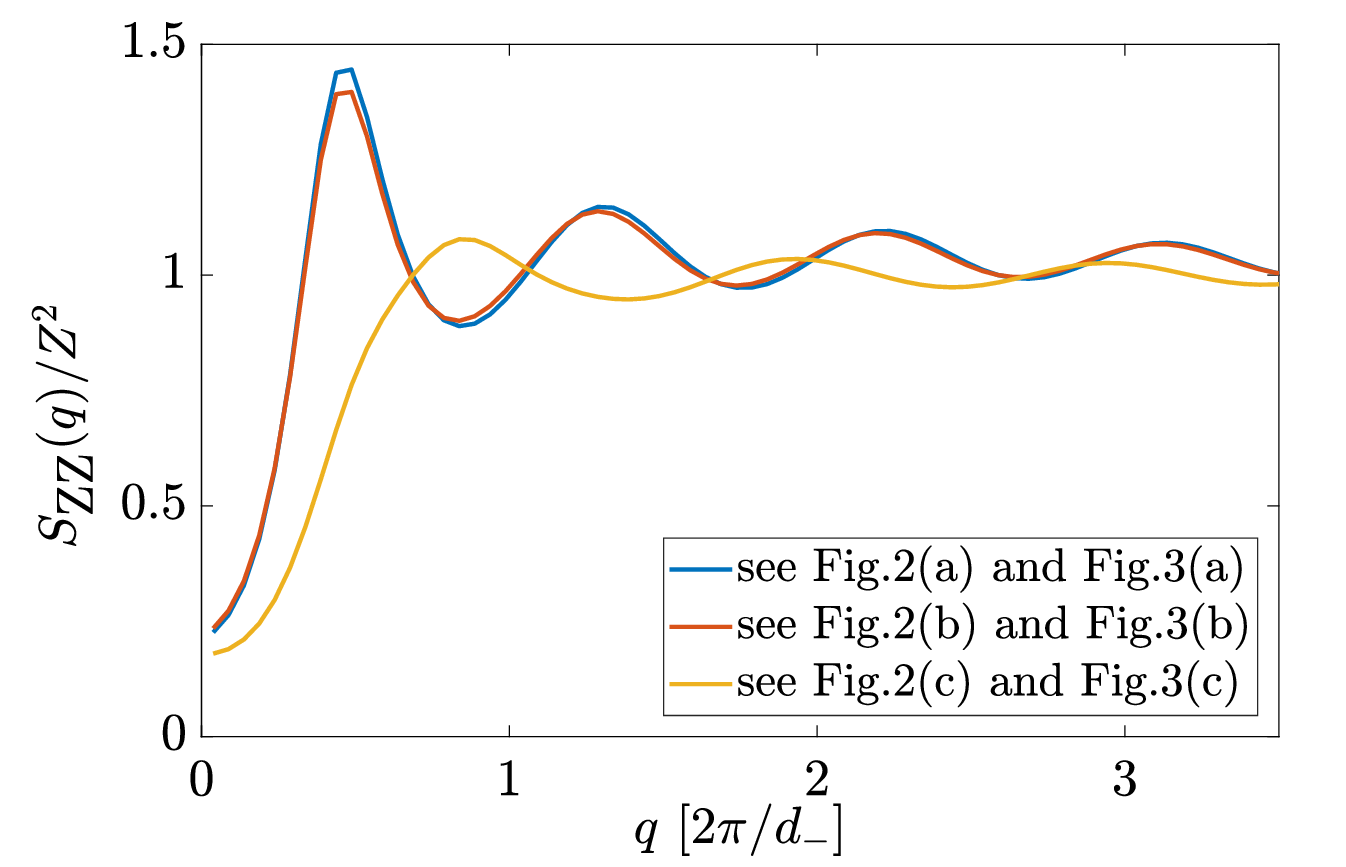}
    \caption{The number-number (NN) and charge-charge (ZZ) structure factor calculated via Eqs.~\eqref{Eq:S_NN} and~\eqref{Eq:S_ZZ}, respectively, corresponding to the structure factors shown in Fig.~\ref{Fig:rho}, and obtained from MSAc for the system parameters given in Figs.~\ref{Fig:rho} and~\ref{Fig:S}.}
    \label{Fig:S_NN_ZZ}
\end{figure}

Let us then consider the structure factors, which are calculated in MD \cite{Hartel_2016} via the definition
\begin{align}\label{Eq:S_MD}
    S_{ij}^{\textrm{MD}}(\mathbf{q};z_0,L') = \left\langle \frac{1}{N_{\textrm{tot}}}\sum_{n\in\Gamma_i}\sum_{m\in\Gamma_j} \exp[-i\mathbf{q}\cdot (\rb_n-\rb_m)]\right\rangle,
\end{align}
where $N_{\mathrm{tot}}$ denotes the total number of particles within the slab $[z_0-L'/2,z_0+L'/2]$, and $\Gamma_i$  the set of particles of species $i$ within the slab. We define $\mathbf{q}=(q_x,q_y,0)$ as the in-plane wave vector with $q=|\mathbf{q}|$ the radial wave number defined before.
Eq.~\eqref{Eq:S_MD} is comparable to the structure factor calculated from Eq.~\eqref{Eq:S_ipin}. 
To demonstrate this, we compare 
results from MD using Eq.~\eqref{Eq:S_MD} (circles) with MSAc using Eq.~\eqref{Eq:H} (solid lines)  in Fig.~\ref{Fig:S}. 
For this purpose, we define, in accordance with Eq.~\eqref{Eq:S_ipin}, $H_{ij}^{\textrm{MD}}=\tfrac{N_\textrm{tot}}{N_i} S_{ij}^{\textrm{MD}}$. 
The blue lines depict the $H_{++}$ component, the orange line the $H_{--}$ component, and the yellow line the $H_{+-}$ component. Note that we cannot show the $H_{-+}$ values for MD, because it was omitted in Ref.~\onlinecite{Hartel_2016}, due to poor statistics. The MSAc functional performs a lot better than the MFC functional 
(not shown here, but can be found in Ref.~\onlinecite{Hartel_2016}), and the agreement between MD and MSAc is satisfying. Since MD did not provide the $S_{-+}$ components due to poor statistics, we cannot properly compare the $S_\textrm{ZZ}$ and $S_\textrm{NN}$ against simulations. However, given the agreement between DFT and MD, as seen in Fig.~\ref{Fig:S}, we presume satisfying agreement for $S_\textrm{ZZ}$ and $S_\textrm{NN}$ as well. The corresponding charge-charge and number-number structure factors $S_\textrm{ZZ}$ and $S_\textrm{NN}$ from DFT, respectively, are plotted in Fig.~\ref{Fig:S_NN_ZZ}. Note the shift in the first maximum in $S_\textrm{ZZ}$, which will be discussed in the following section~\ref{sec:structure-results}. 

\section{The Structure Changes with the Surface Potential}\label{sec:structure-results}
\begin{figure*}
    \centering
    \includegraphics[width=\columnwidth]{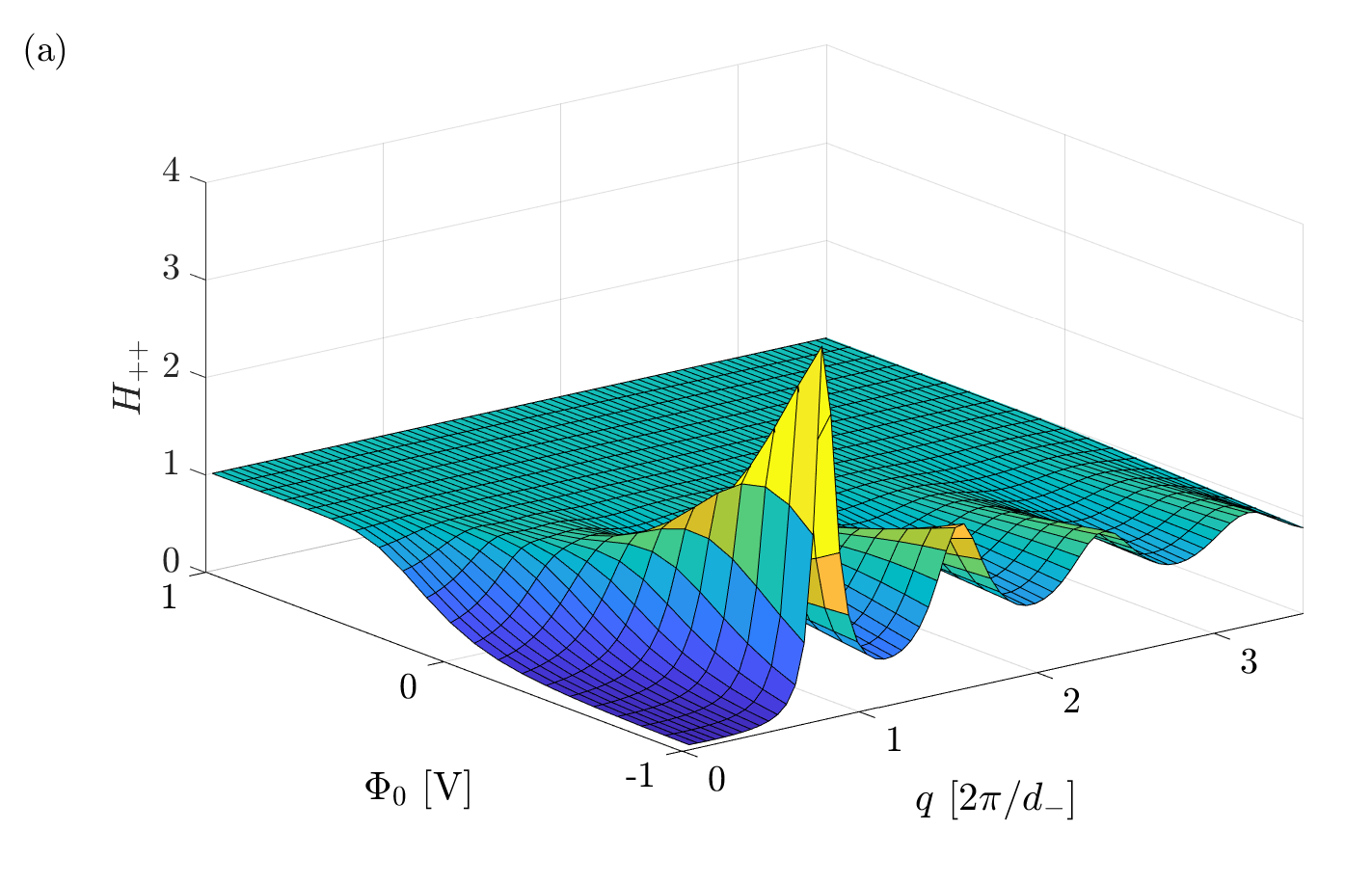}
    \includegraphics[width=\columnwidth]{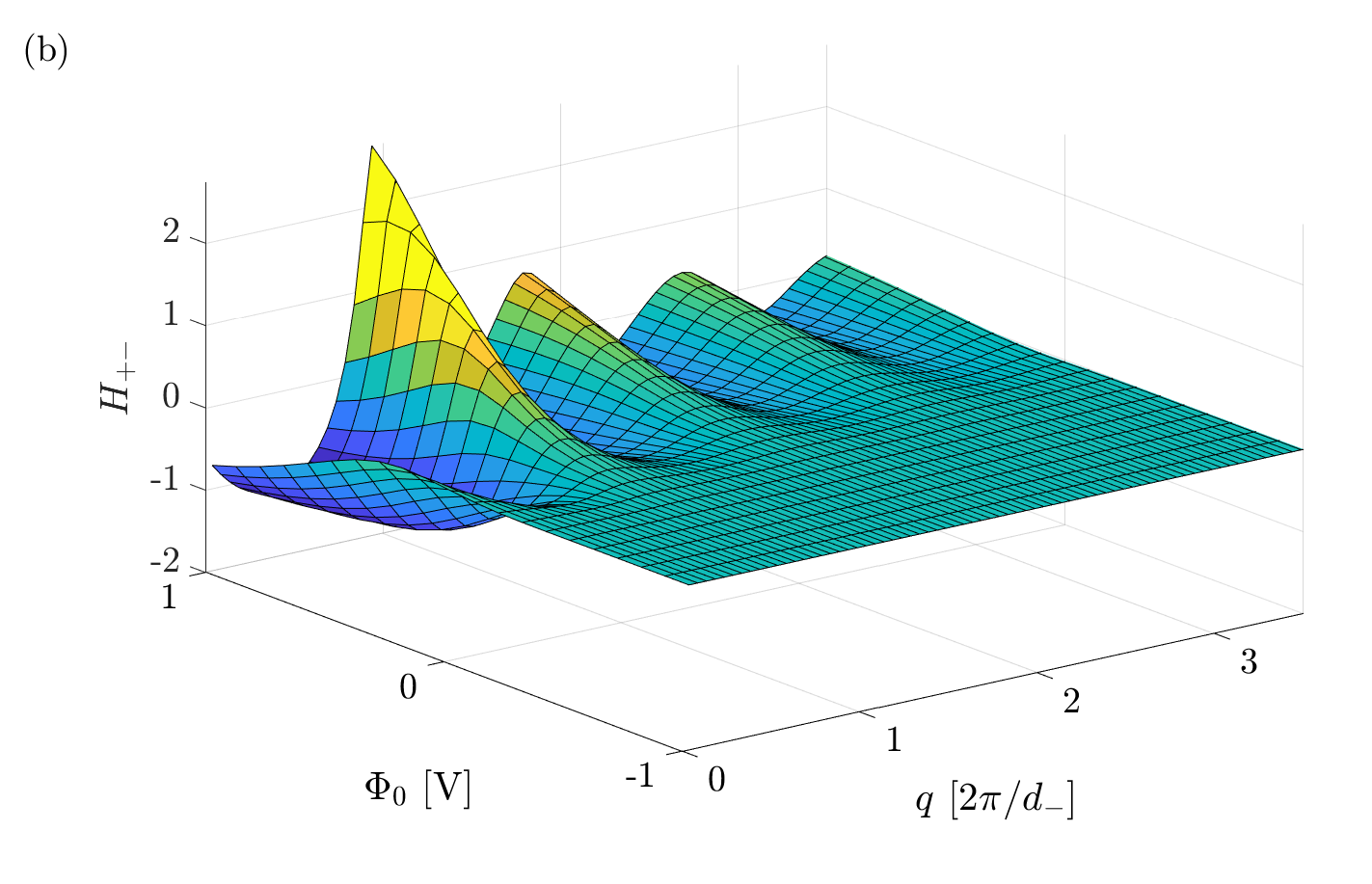}\\
    \includegraphics[width=\columnwidth]{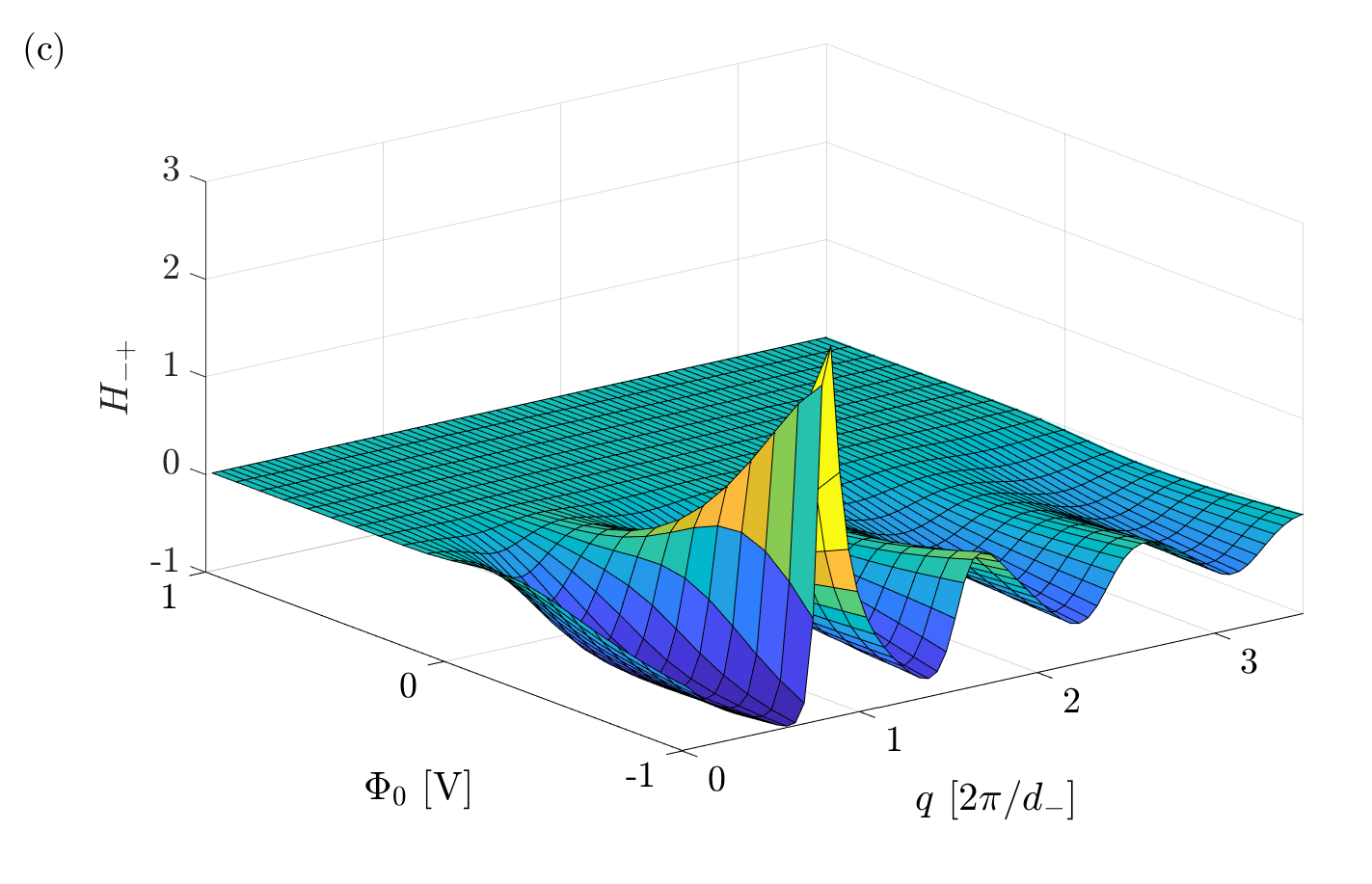}
    \includegraphics[width=\columnwidth]{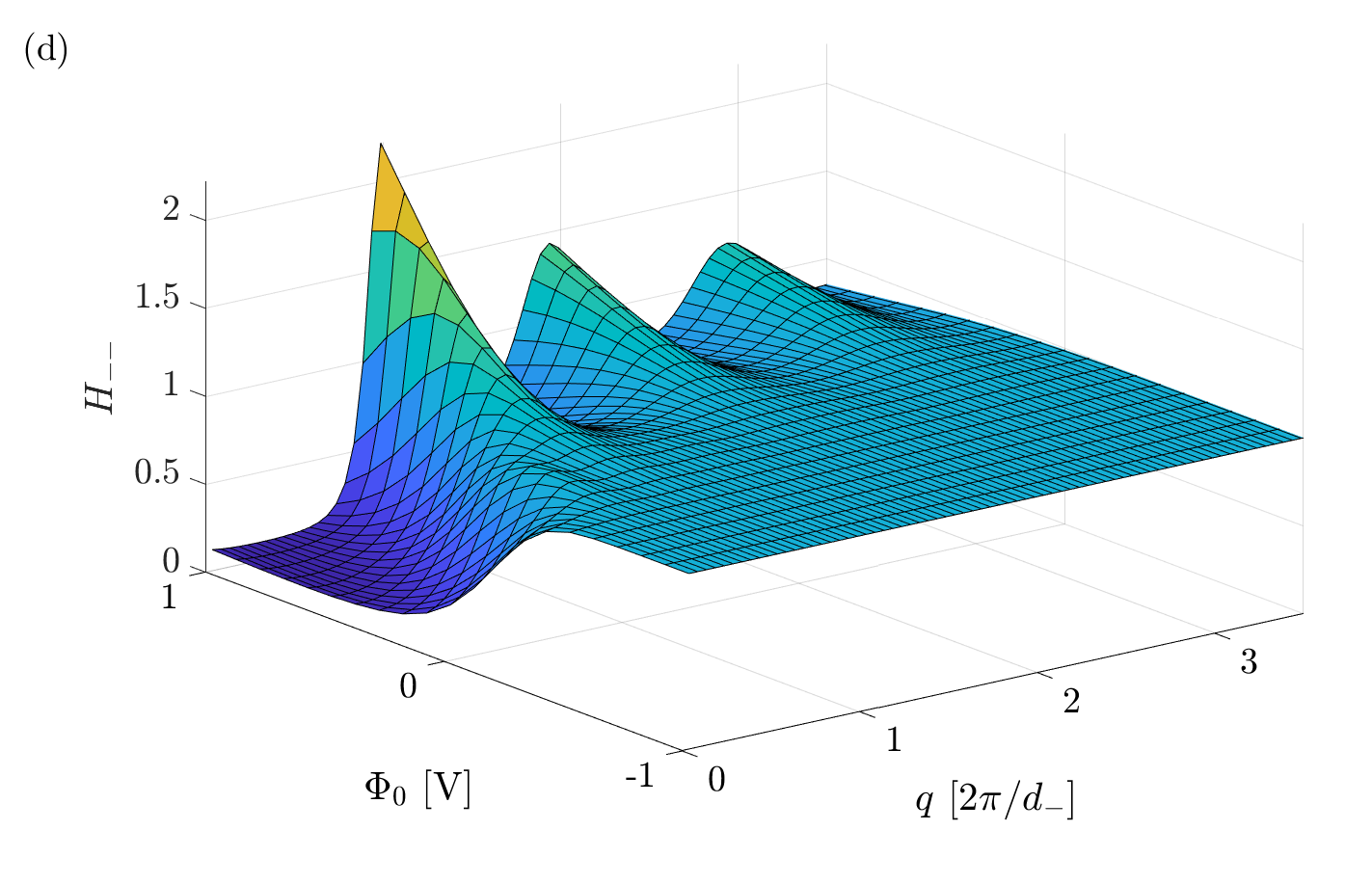}
    \caption{(a)-(d) The normalized structure factor $H_{ij}(q)$ for varying surface potential $\Phi_0$, obtained via Eq.~(\ref{Eq:H}) from DFT using the MSAc functional. Note the different color schemes in each panel.}
    \label{fig:H_V}
\end{figure*}

\begin{figure}
\includegraphics[width=\columnwidth]{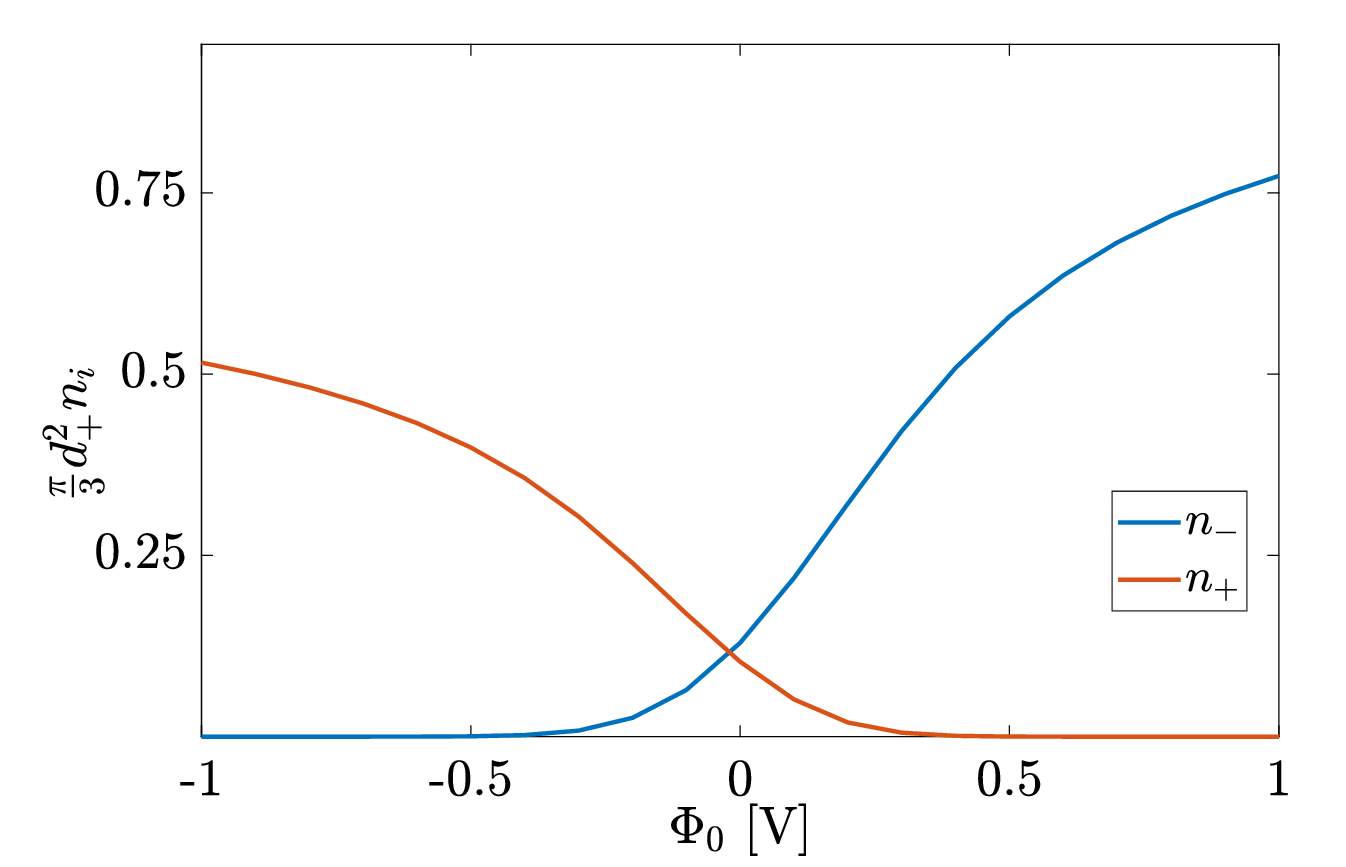}
\caption{\label{Fig:n} The average packing fraction $\tfrac{\pi}{6}\frac{n_i(z_0,L')}{L'}d_+^3$ within the first layer to the electrode with $L'=d_+/2$. }
\end{figure}

Now we have a well-performing formalism at hand to investigate the structure factor as a function of the surface potential $\Phi_0$. In Fig.~\ref{fig:H_V}(a)-(d) we show the normalized in-plane structure factors $H_{ij}(q)$ for varying surface potential $\Phi_0$. The surface plots show $H_{ij}(q)$ on the $z$-axis, the wave number $q$ on the $x$-axis, and the surface potential $\Phi_0$ on the $y$-axis. By plotting $H_{ij}$, one emphasizes the role that the cat- and anions play at a given surface potential; at positive potentials $\Phi_0\gtrsim 0.5$ V the (negative) anions dominate (see $H_{+-}$ and $H_{--}$), while at negative potentials $\Phi_0\lesssim -0.5$ V the (positive) cations dominate (see $H_{-+}$ and $H_{++}$). To exemplify this, we plotted the average packing fraction of the cations and anions in the first layer to the electrode in Fig.~\ref{Fig:n} as function of the applied surface potential $\Phi_0$.

\begin{figure*}
    \includegraphics[width=\columnwidth]{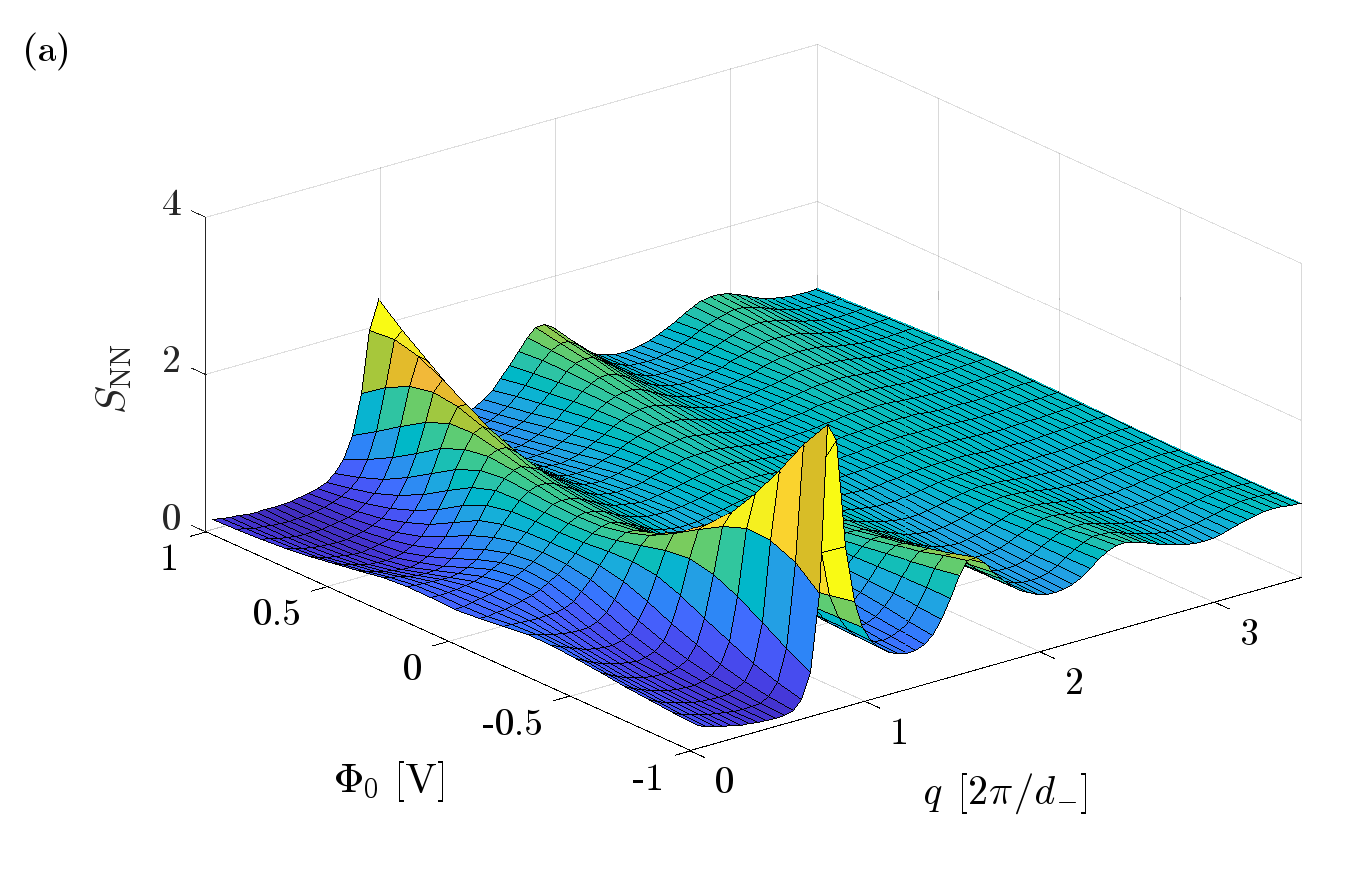}
    \includegraphics[width=\columnwidth]{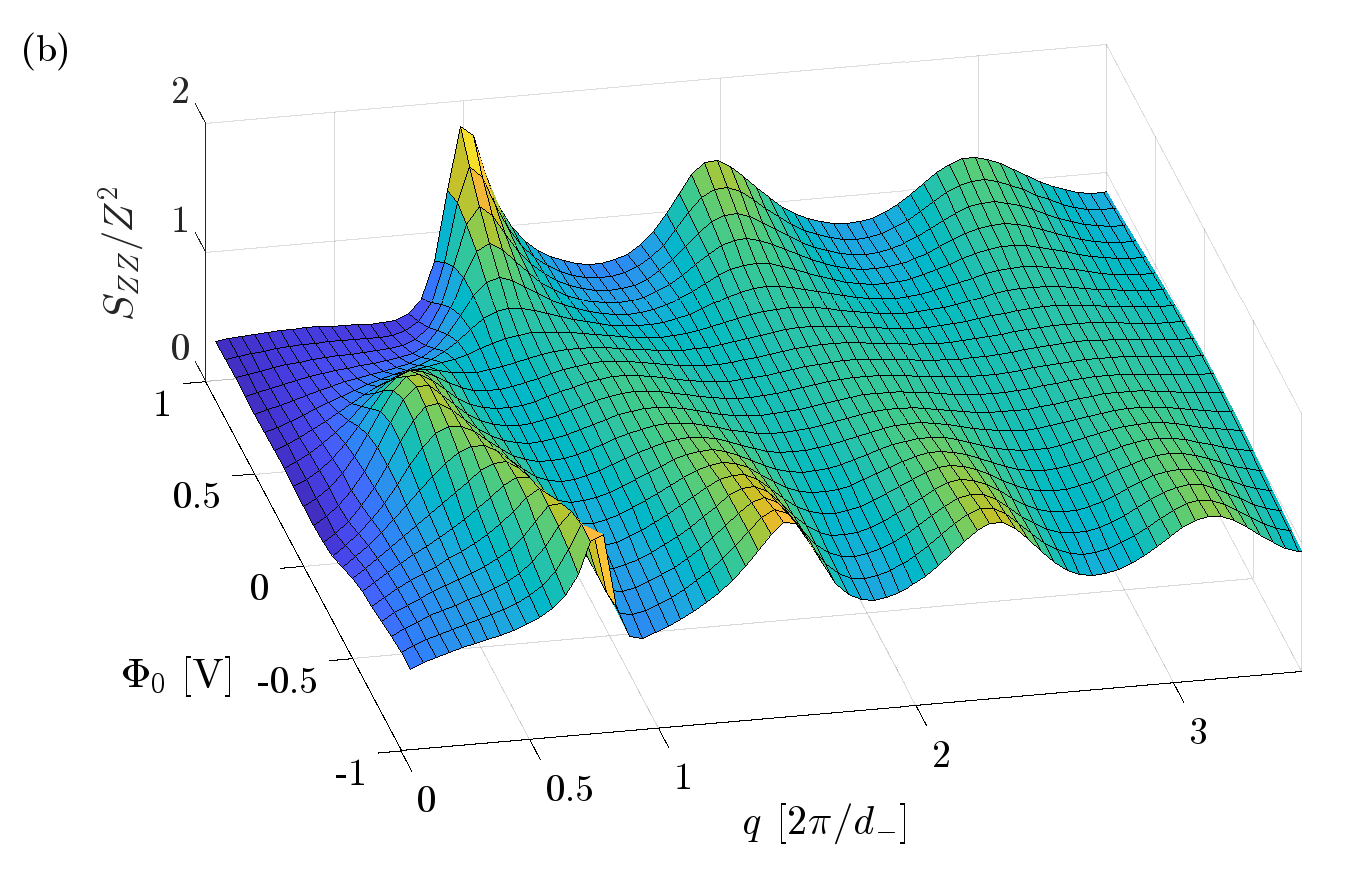}\\
    \caption{    
    The number-number (NN) and charge-charge (ZZ) structure factor calculated via Eqs.~\eqref{Eq:S_NN} and~\eqref{Eq:S_ZZ}, respectively, corresponding to the structure factors shown in Fig.~\ref{Fig:rho}, and obtained from MSAc for the system parameters given in Figs.~\ref{Fig:rho} and~\ref{Fig:S}. Note that in (b) $S_{\mathrm{ZZ}}/Z^2$ is shown with $Z^2=0.785^2\approx 0.62$.
    }
    \label{fig:S_V}
\end{figure*}

In Ref.~\onlinecite{Merlet_2014}, the authors found an ordered phase at small surface potentials resulting in large values for the peak height of the structure factor. As mentioned before, we cannot find this ordered structure directly due to the construction of our theoretical approach. Nevertheless, we would expect to find precursors of an ordering transition, if it exists. In Fig.~\ref{fig:S_V}(a) and (b) we plot the number-number and charge-charge structure factor $S_\textrm{NN}$ and $S_\textrm{ZZ}$, respectively, as function of the surface potential $\Phi_0$ and wave number $q$. Similar as in Ref.~\onlinecite{Merlet_2014}, there is a small bumb in $S_\textrm{ZZ}$ at small surface potentials ($q\approx\pi/d_{-}$). Interestingly, the location of this maximum (or bump) in Fig.~\ref{fig:S_V}(b) is located around $q=0.45*2\pi/d_- = 2\pi/(d_-+d_+)$ for vanishing surface potentials, and shifts to $q=2\pi/d_-$ and $q=0.81*2\pi/d_-=2\pi/d_+$ at positive and negative potentials, respectively. The reason being that at small surface potentials, the difference in the number of cations and anions within the first layer is small  (see Fig.\ref{Fig:rho}(a)) and therefore, within the plane, each cation (anion) is approximately surrounded by an anion (cation) and the charge-charge structure spans two ion diameters resulting in a peak at $q=2\pi/(d_-+d_+)$.   At larger (absolute) surface potentials, the first layer gets fully filled with counterions, and therefore the charge-charge structure couples to the number-number structure, causing a peak in $S_\textrm{ZZ}$ at the inverse ion diameter $2\pi/d_j$. Hence, we do find a structural change in $S_\textrm{ZZ}$ in the first layer near the electrode from a diffuse EDL at small surface potentials to a dense-packed EDL at larger surface potentials. 

\begin{figure*}
\includegraphics[width=\columnwidth]{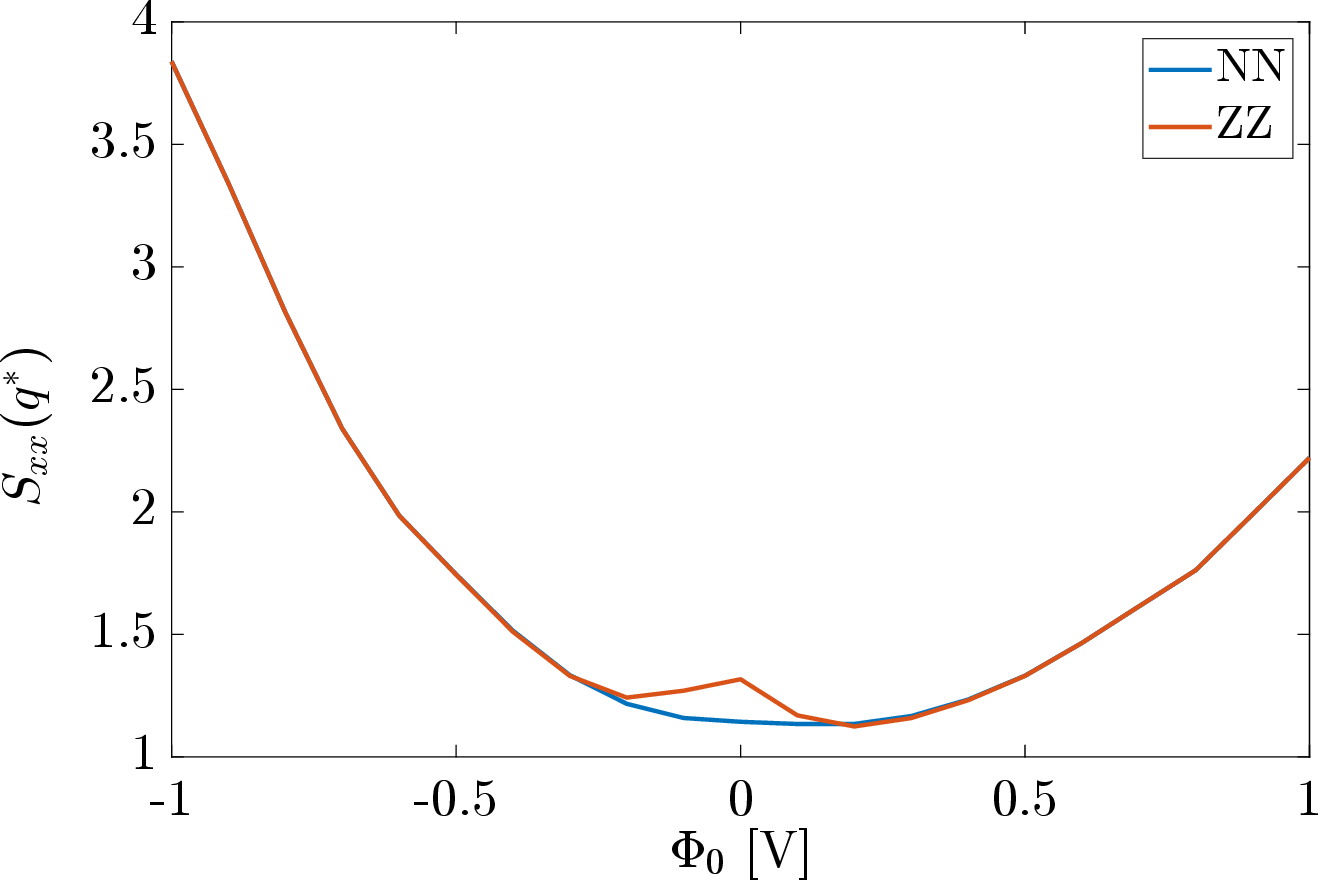}
\caption{\label{Fig:max_S} The value of $S_{\mathrm{NN}}$ and $S_{\mathrm{ZZ}}$ as presented in Fig.~\ref{fig:S_V} at its maximum position $q^*$. Note that for a comparison with Fig.~\ref{fig:S_V}(b) one should keep in mind that $Z^2\approx 0.62$.}
\end{figure*}

It might be interesting to compare the first layer of the EDL next to the electrode with a two-dimensional system of hard disks. The size ratio between the positive and negative ions is $d_+/d_-=0.618/0.506\approx 1.22$. For this size ratio, binary hard disks are not expected to form a crystalline-like phase \cite{Huerta_2012}. However, 
Fig.~\ref{Fig:n} shows that at sufficiently large potentials mainly one species of particles fills the first layer to the electrode. 
At large absolute potentials the system exceeds the packing fraction where freezing would be expected: For monodisperse hard disks in two dimensions this is $\eta^{2\textrm{d}}=0.68$ \cite{Huerta_2006}, which would correspond to $\eta\approx 0.46$ for an ideal layer of hard spheres where all spheres are located exactly in the center of the slab of size $L'=d_+/2$. 
This is further reflected in Fig.~\ref{Fig:max_S}, where the height of the first peak $S_{ij}(q^*)$ is shown for both $S_{\textrm{NN}}$ and $S_{\textrm{SS}}$ from Fig.~\ref{fig:S_V}. At sufficiently large potentials $\Phi_0$ the peak exceeds values of $2.8-3.1$, which, for hard spheres in bulk, typically happens close to the freezing transition~\cite{Ramakrishnan_1979}. 
This can be seen as a precursor of an ordering transition as it would be expected at high packing fractions. 
%Note again that in our However, in our approach we assume cylindrical symmetry and, thus, can only see such precursors of freezing but not the transition itself. 

%$\eta^{3d}=\frac{N \pi/6 d^3}{A d}$
%$\eta^{2d}=\frac{N \pi/4 d^2}{A}$
%$\eta^{3d}=\frac{2}{3}\eta^{2d}\approx 0.46$

%\Andreas{[At this point, plotting the in-plane pair-distribution function would be worth, including a few words on the precursor of nucleation (as mentioned at the end of sec. II).]}\PC{[check refs. which one explains crystallization for binary mixtures?]}

\section{Conclusion}\label{sec:conc}
In continuation of previous work \cite{Merlet_2014,Hartel_2016}, we investigated the in-plane structure of EDLs at charged electrodes.
We properly derived the equation for the in-plane structure factor; we used a slab of a certain thickness in which the structure factor is calculated. In comparison to previous work, satisfying agreement was found within the primitive model of electrolytes (PM) between MD simulations \cite{Hartel_2016} and our approach, where we made use of DFT with the MSAc functional. This allowed us to further explore the features of the in-plane structure factor as a function of the electrostatic surface potential. At positive potentials the structure factor is dominated by the anions, while at negative potentials it is dominated by the cations. This finding is in line with expectations and with previous work from Ref.~\onlinecite{Cats_2021} in which the same trend was shown in the differential capacitance. Less trivially, we found that for the charge-charge structure factor $S_\textrm{ZZ}$ at low surface potentials, where neither the cations nor anions dominate, the location of the maximum of the structure factor is located around $2\pi/(d_-+d_+)$, \textit{i.e.} each coion is surrounded by a counterion. At larger surface potentials, however, the location of the maximum converges to $2\pi/d_j$, \textit{i.e.} the first layer is filled with only counterions. This clearly demonstrates that one is able to distinguish between a diffuse EDL and a dense-packed EDL, using our approach. 

The primitive model that we studied here has its known shortcomings. One of the crucial simplifications of the model is considering an implicit solvent. The reason hereof is simplicity, as the solvent molecules, in particular water molecules, have extremely complex interactions and are therefore hard to model accurately. One implication of considering an implicit solvent in our case is that the dielectric permittivity is constant throughout the system, which is certainly not true for real electrolytes near charged surfaces where solvent molecules (and ions to some extend) have the tendency to orient along the electric field~\cite{Marcovitz_2015,Gongadze_2018}. However, there is not a simple theory to take this effect into account. In this manuscript, we want to keep the model as simple as possible while still capturing important features of the real system. Note that the simulations by Merlet et.al~\cite{Merlet_2014} do not have a solvent. Another simplification in our model is to not consider explicitly an inhomogeneous surface-charge-density distribution. Doing so would imply us to invoke three-dimensional DFT calculations, which would defeat the purpose of doing DFT in the first place, as three-dimensional DFT is computationally very expensive.

Overall, we explored the EDL from a novel perspective and presented how to get access to the in-plane structure within the framework of DFT. Although the in-plane structure factor is a tedious object, we showed that it can provide interesting insight in the EDL. One of the most prominent features of the bulk structure factor, which we have not yet mentioned, is that it is a measurable quantity. The main question, therefore, is whether the in-plane structure factor studied in this manuscript theoretically, is also measurable. If so, we established a direct connection between the density profiles, which are very hard (if not impossible) to measure~\cite{Chu_2023}, to the measurable in-plane structure factor. This would open a doorway for a much more thorough understanding of the EDL. 

\section*{Acknowldegments}
We acknowledge funding from the German Research Foundation (DFG) through Project Number 406121234. We acknowledge support by the state of Baden-Württemberg through bwHPC and the German Research Foundation (DFG) through Grant No. INST 39/963-1 FUGG
(bwForCluster NEMO).

\section*{Data Availability}
The data that support the findings of this study are available from the corresponding author
upon reasonable request.

\bibliography{apssamp}% Produces the bibliography via BibTeX.

\appendix

%%%%%%%%%%%%%%%%%%%%%%%%%%%%%%%%%%%%%%%%%%%%%%%%%%%%%%%%%%%%%%%
\section{Mean Spherical Approximation} \label{App:MSA}
%%%%%%%%%%%%%%%%%%%%%%%%%%%%%%%%%%%%%%%%%%%%%%%%%%%%%%%%%%%%%%%
%\PC{For completeness, the following section is from Ref.. Rewrite a bit.}

The grand potential that lies at the basis of DFT is given by
\begin{align}
\Om = &\F_\textrm{id}[\rhos] + \F_\textrm{ex}[\rhos]-\nonumber\\
&\sum_j \int \dd\rb \rho_j(\rb) (\mu_j-V_\textrm{ext}^j(\rb)),
\end{align}
where $\F_\textrm{id}[\rhos]$ denotes the ideal Helmholtz free energy function of the non-interacting system, $\F_\textrm{ex}[\rhos]$ the excess (or over-ideal) Helmholtz free energy functional that takes into account the interparticle interactions, $\mu_j$ the chemical potential, and $V_\textrm{ext}^j(\rb)$ the external potential of species $j$. The excess functional $\F_\textrm{ex}[\rhos]$ that is used throughout this work reads
\begin{align}
\F_\textrm{ex}[\rhos] = \F_\textrm{ex}^\textrm{HS}[\rhos]+\F_\textrm{ex}^\textrm{MFC}[\rhos]+\F_\textrm{ex}^\textrm{MSAc}[\rhos],
\end{align}
where $\F_\textrm{ex}^\textrm{HS}[\rhos]$ is the White-Bear II implementation of Fundamental Measure Theory dealing with the steric hard-sphere interactions (e.g. see Refs.~\onlinecite{Roth_2010,Hansen_Goos_2006}), $\F_\textrm{ex}^\textrm{MFC}[\rhos]$ the mean-field functional dealing with Coulombic interactions, and $\F_\textrm{ex}^\textrm{MSAc}[\rhos]$ the corrections on top of MFC using results from the Mean-Spherical-Approximation~\cite{Waisman_1970,Waisman_1972_I,Waisman_1972_II,Blum_1,Blum_2,Blum_3,Blum_c,Hiroike,MSAc}. For further details and performance of this functional we refer to previous work~\cite{Voukadinova_2018,Cats_2020_decay,Cats_2020}. One of the strengths of DFT is that one can consider any external potential without increasing the complexity of the calculations. For now we consider charged hard walls, for which the non-electrostatic part of the external potential in a planar geometry reads
\begin{align}
\beta V_\textrm{ext}^{j}(x) = \begin{cases}
\infty &\quad \text{ for } x \leq d_{j}/2;\\
0&\quad \text{ for } x > d_{j}/2,\\
\end{cases}
\end{align}
where $x$ is the coordinate perpendicular to the electrode and $d_j$ the hard-sphere diameter of particle species $j$.

One can obtain the direct correlation function from the excess functional according to
\begin{align}
    c^{(2)}_{ij}(\rb,\rb') = -\beta \frac{\delta^2 \F_\textrm{ex} [\rhos]}{\delta \rho_i(\rb)\delta \rho_j(\rb')}.
\end{align}
By assuming translational symmetry in the $xy$-plane, one can write the direct correlation function as $c^{(2)}_{ij}(\rb,\rb') = c^{(2)}_{ij}(z,z',u)$, where $u$ denotes the radial component in the $xy$-plane in cylindrical coordinates (distance between particles in the plane projection).

\subsection{The excess functional}

Now, the electrostatic mean-field functional is given by 
\begin{align}\label{Eq:F_ex_meanfield}
  \F_\textrm{ex}^\textrm{MFC}[\{\rho\}] =&A\frac{1}{2}\int \dd z \rho_\textrm{Z}(z)\Phi(z) , 
\end{align}
an integration over the average Coulombic energy per position. 
Its related direct-correlation functions are $c_{ij}^\textrm{MFC}$. 
Then, $\Delta c_{ij}^\textrm{MSAc}(z)$ is a correction on top of mean-field theory based on the mean spherical approximation (MSA). 
The related Helmholtz free energy functional is given by 
\begin{align}\label{Eq:F_ex_MSAc}
 \F_\textrm{ex}^\textrm{MSAc}[\{\rho\}]=&-A\frac{k_\textrm{B}T}{2}\sum_{ij}\int\dd z\int\dd z' \rho_i(z)\nonumber\\
     &\quad \quad \times \Delta c^\textrm{MSAc}_{ij}(|z-z'|;\{\rho_\textrm{b}\})\rho_j(z') . 
\end{align}
In the bulk, $\Delta c^\textrm{MSAc}(r)$ is given by~\cite{Hiroike}
\begin{align}\label{Eq:app_cMSAc}
    \Delta c_{ij}^\textrm{MSAc}(r)=
    \left\{
    \begin{matrix}
     c_{ij}^\textrm{MSAsh}(r)+z_i z_j\frac{\lambda_\textrm{B}}{r}, & r<\Delta d_{ij}; \\
     c_{ij}^\textrm{MSAl}(r)+z_i z_j\frac{\lambda_\textrm{B}}{r}, 
     & \Delta d_{ij}\leq r<d_{ij};\\
     0, 
     & d_{ij}\leq r;
    \end{matrix}
    \right.
\end{align}
with $\Delta d_{ij}=|d_i-d_j|/2$ and $d_{ij}=(d_i+d_j)/2$. The term for short-ranged (sh) correlations with $d_i<d_j$ reads 
\begin{align}\label{Eq:app_cMSAsh}
    c_{ij}^\textrm{MSAsh}(r)=2\lambda_\textrm{B}\left(z_i N_j+d_i\zeta(X_i+\frac{1}{3}d_i^2\zeta)\right),
\end{align}
with the parameters 
\begin{align}
    X_j=&\frac{z_j-d_j^2\zeta}{1+\Gamma d_j}; 
    & 
    N_j=&\frac{X_j-z_j}{d_j}; \notag \\
    \Gamma=&\pi\lambda_\textrm{B}\sum_j\rho_{\textrm{b},j}X_j^2; 
    &
    \zeta=&\frac{1}{H}\sum_j\frac{\rho_{b,j}d_jz_j}{1+\Gamma d_j}; \notag \\
    H=&\sum_j\frac{\rho_{\textrm{b},j}d_j^3}{1+\Gamma d_j}+\frac{2}{\pi}\left(1+\eta_b\right); 
    &
    \eta=&\frac{\pi}{6}\sum_jd_j^3\rho_{\textrm{b},j}. \notag
\end{align}
The term for long-ranged (l) correlations in Eq.~\eqref{Eq:app_cMSAc} with $\Delta d_{ij}<r<d_{ij}$ is given by
\begin{align}
    c_{ij}^\textrm{MSAl}(r)=\frac{\lambda_\textrm{B}}{r}\left[A_{ij}+B_{ij}r+C_{ij}r^2+F_{ij}r^4\right] ,\label{Eq:app_cMSAl}
\end{align}
where
\begin{align}
  A_{ij} =& -\Delta d_{ij}^2\left[\zeta(X_i+X_j)+\zeta^2\Delta d_{ij}^2-N_iN_j\right],\\
  B_{ij}=&-(X_i-X_j)(N_i-N_j)-(X_i^2+X_j^2)\Gamma-\nonumber\\
         &2d_{ij}N_iN_j+\frac{1}{3}\zeta^2(d_i^3+d_j^3),\\
  C_{ij}=&-\zeta(X_i+X_j)+N_iN_j-\frac{1}{2}\zeta^2(d_i^2+d_j^2),\\
  F_{ij}=&\frac{1}{3}\zeta^2.
\end{align}

\section{The in-plane Structure Factor Moving through the System}\label{sec:sys2}

If one wants to study the two-dimensional structure factor at a position $z_0$, one should apply a dimensional crossover, \textit{i.e.} $\rho_i(z_0)=\left.\rho_i^{(2\textrm{D})}\right|_{z=z_0}\delta(z-z_0)$. This would result in 
\begin{align}
    S_{ij}^{(2\textrm{D})}(q,z_0) = & \frac{\left.\rho_i^{(2\textrm{D})}\right|_{z=z_0}}{\left.\rho_{\mathrm{tot}}^{(2\textrm{D})}\right|_{z=z_0}}\Bigl[\delta_{ij}+\left.\rho_j^{(2\textrm{D})}\right|_{z=z_0}\hat{h}_{ij}(z_0,q)\Bigr],
\end{align}
which is simply the structure factor for a two-dimensional system with surface densities $\rho_j^{(2\textrm{D})}$. Inconveniently, however, $\left.\rho_i^{(2\textrm{D})}\right|_{z=z_0}$ is not well defined in a three-dimensional system, because the probability of finding the centre of a particle exactly at a certain position $z_0$ vanishes. In order to remedy this inconvenience, one can consider weighted densities like $n_i(z_0,L')$, for instance, with small $L'$, which would result in
\begin{align}\label{Eq:S_2D_dirty}
    S_{ij}^{\approx(2\textrm{D})}(q,z_0,L') = \frac{n_i(z_0,L')}{n_{\mathrm{tot}}(z_0,L')}H_{ij}(q,z_0,L')\\
    H_{ij}^{\approx(2\textrm{D})}(q,z_0,L')=\delta_{ij}+n_j(z_0,L')\hat{h}_{ij}(z_0,q).\label{Eq:H_2D_dirty}
\end{align}
Note that this is a good approximation as long as $L'$ is sufficiently small such that $\hat{h}_{ij}(z_0,z,q)\approx \hat{h}_{ij}(z_0,z_0,q)\equiv \hat{h}_{ij}(z_0,q)$  for $z\in [z_0-L',z_0+L']$.
Needless to say, this is an ad-hoc approach and one should wonder whether it is an appropriate and relevant one. However, Eq.~\eqref{Eq:S_2D_dirty} turns out to be computationally convenient when studying the change in the structure factor as function of the position in the system, as investigated in section~\ref{sec:sys2}.

\begin{figure*}
    \centering
    \includegraphics[width=\columnwidth]{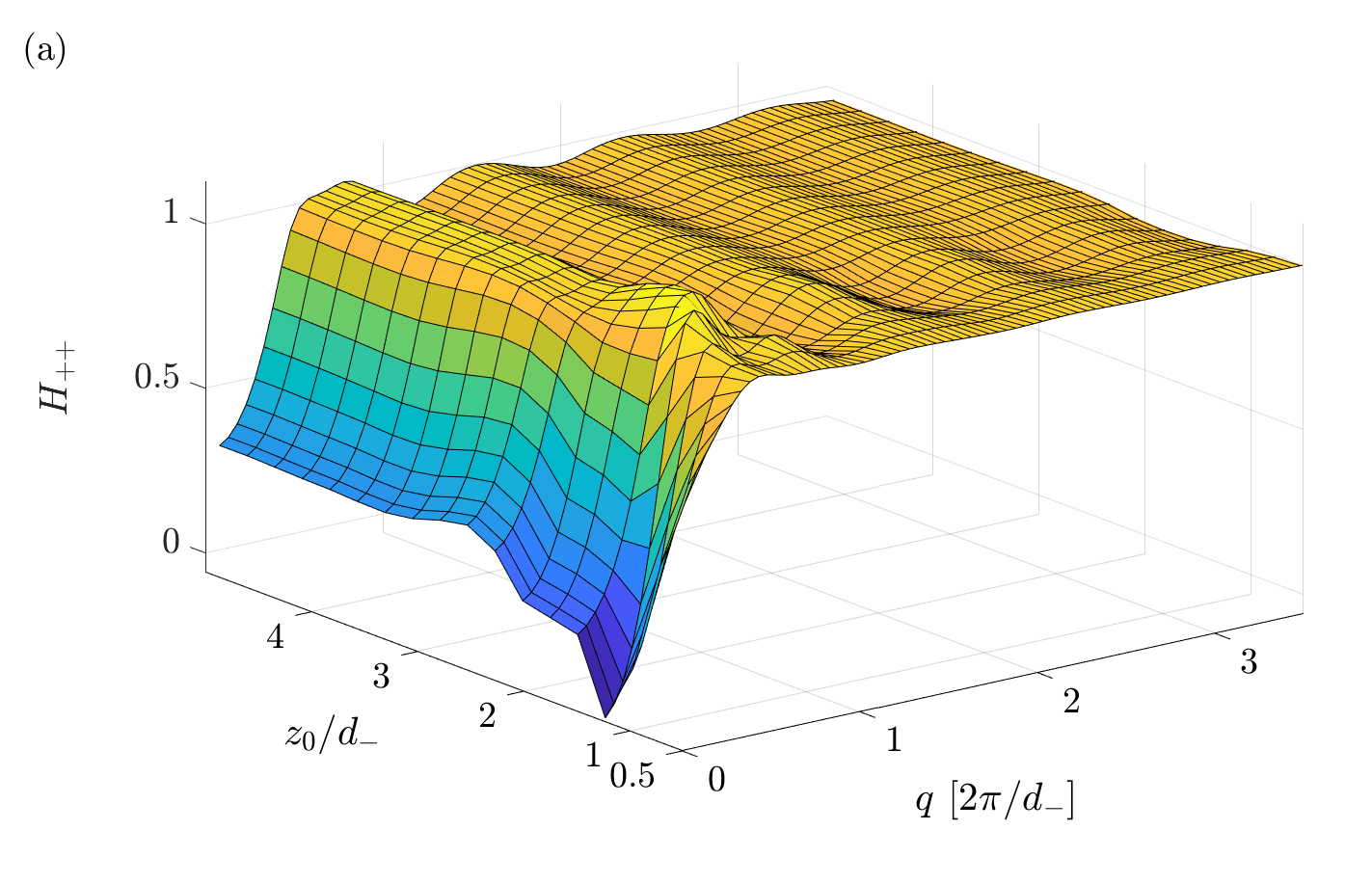}
    \includegraphics[width=\columnwidth]{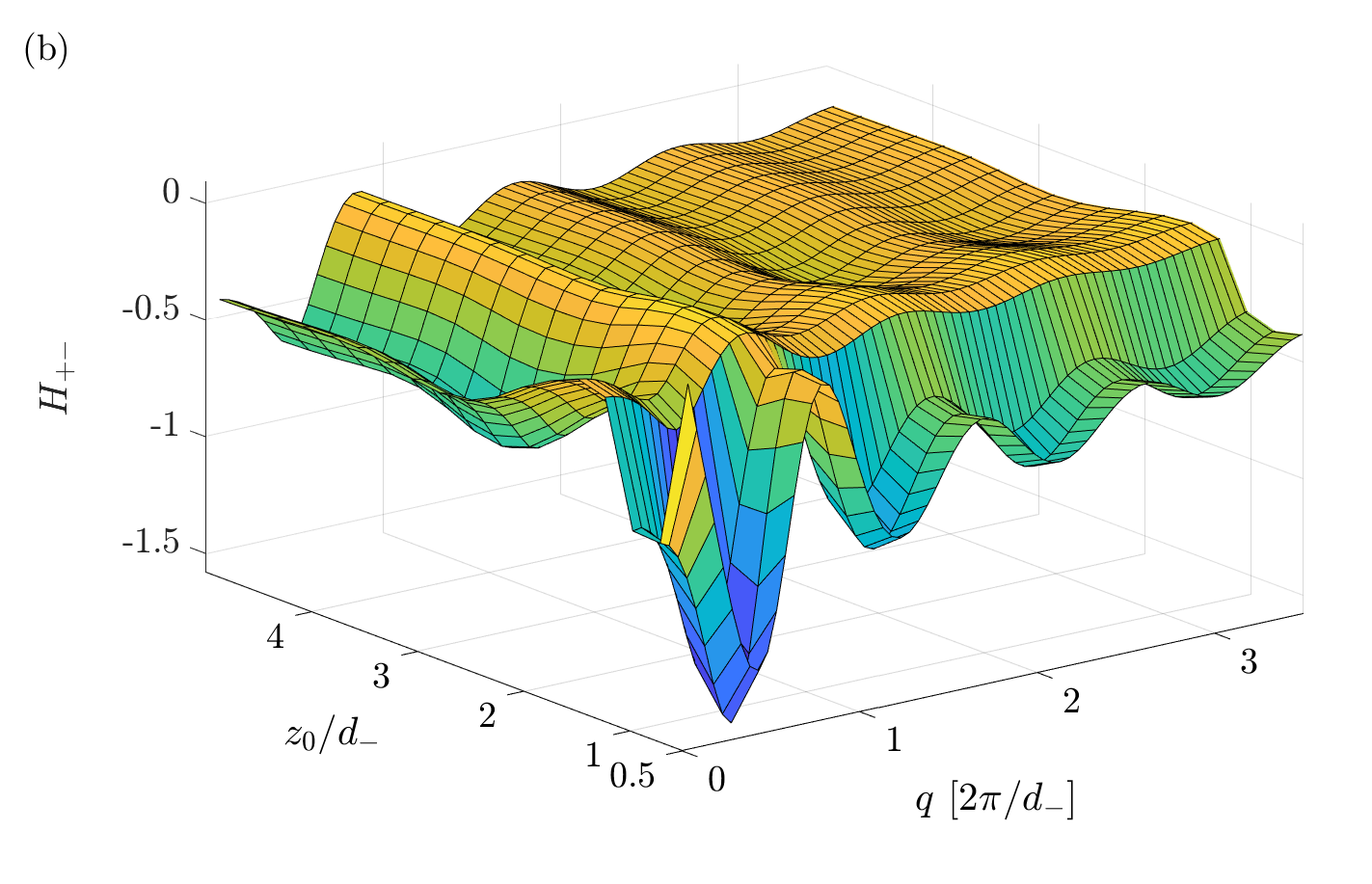}\\
    \includegraphics[width=\columnwidth]{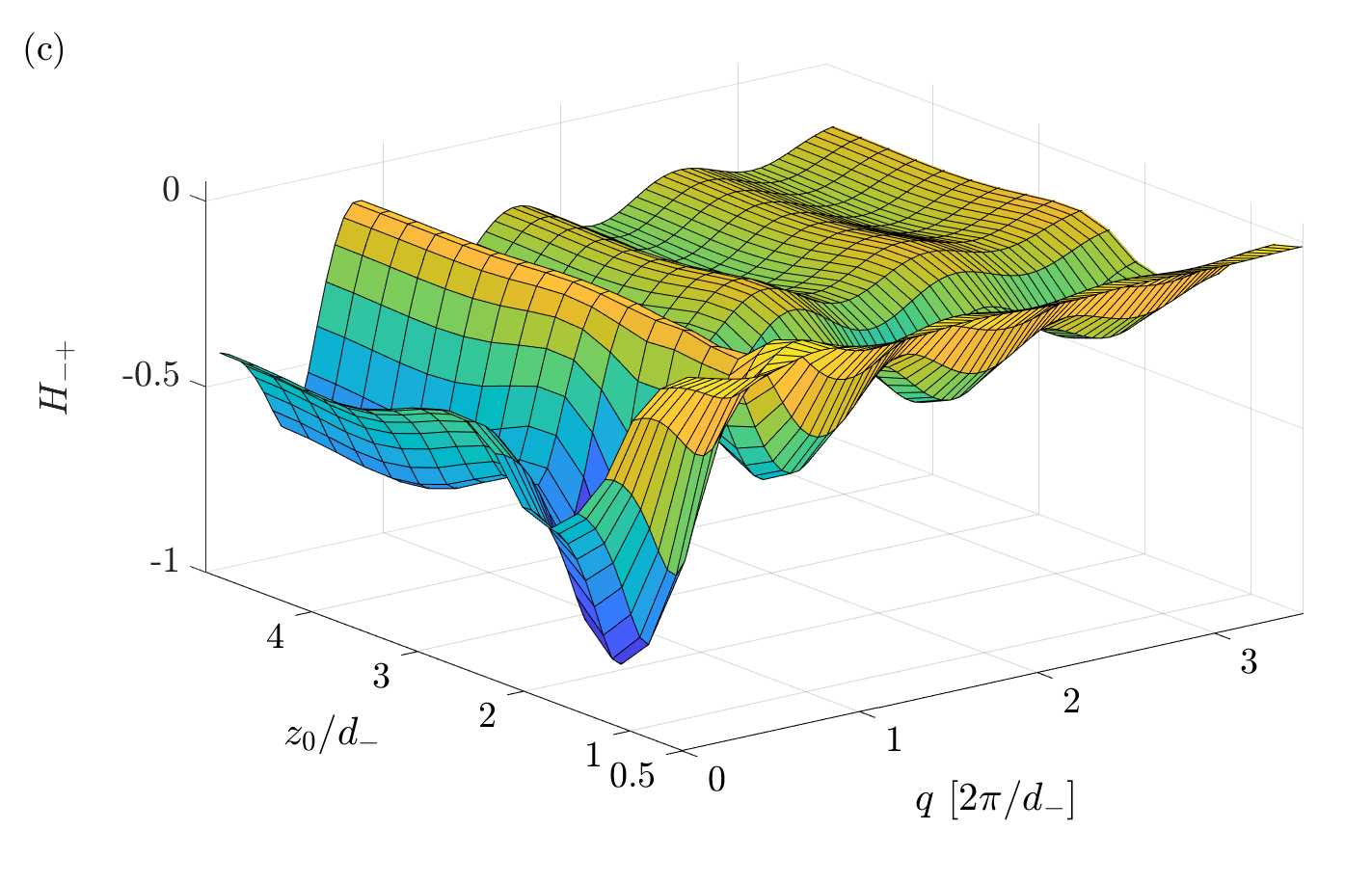}
    \includegraphics[width=\columnwidth]{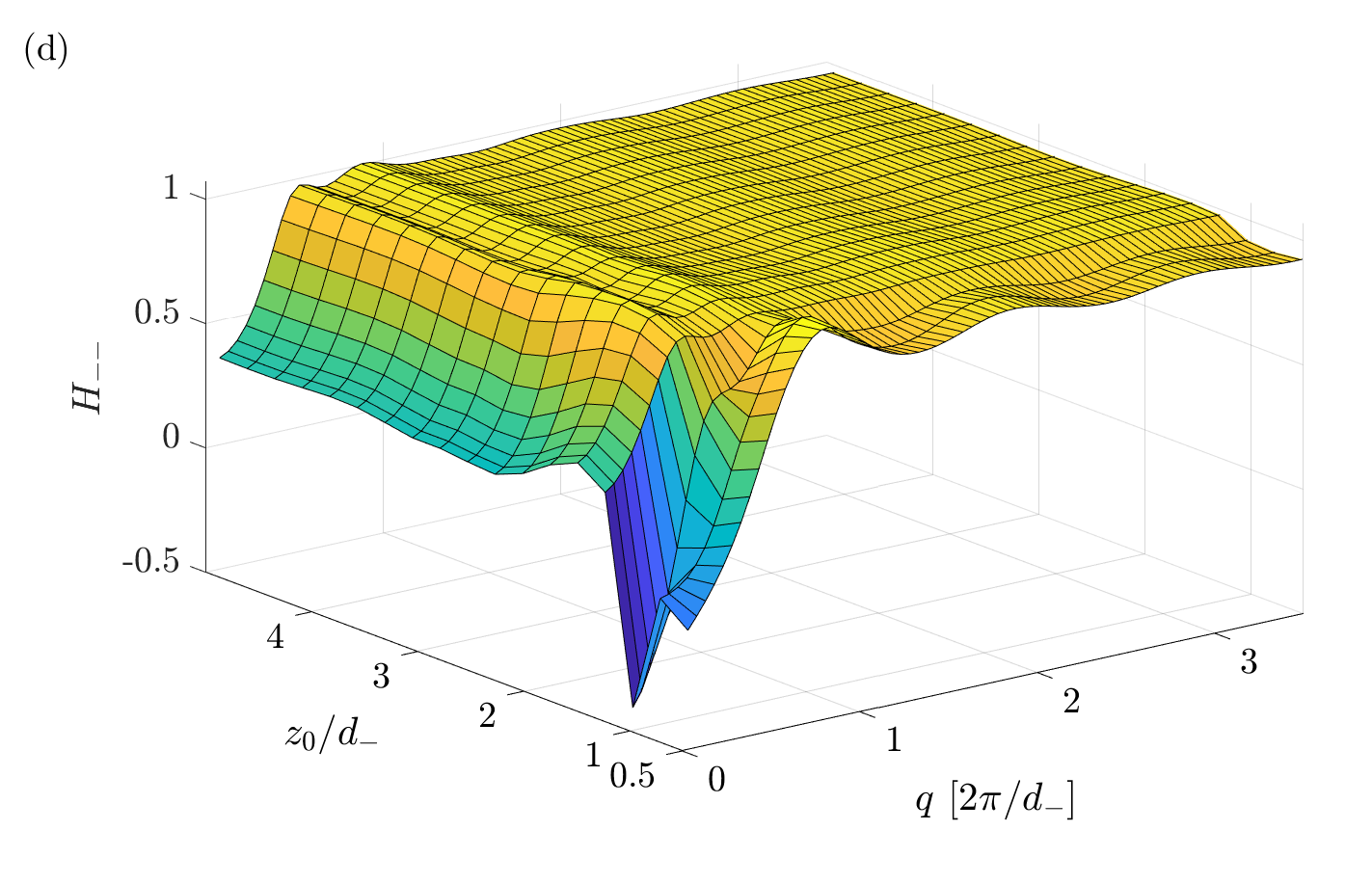}
    \caption{The normalized approximate 2D structure factor as function of $q$ and the position $z$.}
    \label{Fig:H_z}
\end{figure*}

More insight in the structure factor will be found when investigating how it changes as function of the position $z_0$. For this purpose we plot all components of $H_{ij}^{\approx(2\textrm{D})}$ from Eq.~\eqref{Eq:H_2D_dirty} in Fig.~\ref{Fig:H_z} for the system parameters given in Fig.~\ref{Fig:rho}(c). We choose a fixed $L'=d_-/2$, and changed $z_0$ from contact $z_0=d_-/2$ to a bulkish region at $z_0 = 5 d_-$. Interesting to notice is the clear structure in the anion-dominated components $H_{--}$ and $H_{-+}$ for $z_0\lessapprox 1$ and $2 \lessapprox z_0\lessapprox 3$, while the cation-dominated components $H_{++}$ and $H_{+-}$ have more structure for $1 \lessapprox z_0 \lessapprox 2$, which are exactly the regions at which $\rho_-(z)>\rho_+(z)$ and $\rho_+(z)>\rho_-(z)$, respectively (see Fig.~\ref{Fig:rho}). For increasing $z_0$, one finds converging structure factors, which is what one expects since it should converge to its bulk value. However, the use of Eq.~\eqref{Eq:H_2D_dirty} is limited as discussed before, and should be used merely as an indicator for the structure than that it actually presents the structure factor. Nevertheless, we get an understanding on how the in-plane structure $H_{ij}$ of each component $ij$ changes perpendicular to the surface as function  of distance from the surface. 

\end{document}